\documentclass[oneside,12pt]{article}
\usepackage{amsfonts}
\usepackage{epsf,epsfig}
\usepackage{amsmath}
\usepackage{fancybox}
\usepackage{psfig}
\usepackage{amssymb}
\newenvironment{fminipage}%
{\vskip .5em\noindent\begin{Sbox}\begin{minipage}}%
{\end{minipage}\end{Sbox}\fbox{\TheSbox}\vskip .5em}

\newlength{\mylength}
\newcommand{\ket}[1]{\left | \, #1 \right \rangle}
\newcommand{\bra}[1]{\left \langle \, #1 \right |}

\newcommand{\X}{\mathbf{X}}

\newcommand{\x}{\mathbf{x}}
\newcommand{\y}{\mathbf{y}}
\newcommand{\z}{\mathbf{z}}
\newcommand{\s}{\mathbf{s}}

\newcommand{\0}{\mathbf{0}}

\newcommand{\G}{\mathbb{G}}

\renewcommand{\mod}{\hspace{1mm}\textnormal{mod}\hspace{1mm}}

\renewcommand{\Pr}{\text{Pr}}

\newcommand{\GCD}{\text{GCD}}

\newcommand{\poly}{\text{poly}}
\newcommand{\polylog}{\text{polylog}}

\newcommand{\Z}{\mathbb{Z}}

\newcommand{\R}{\mathbb{R}}

\makeindex

\begin{document}

\titlepage
\begin{center}
Quantum Algorithms

\vspace{5mm}

{Michele Mosca \\
Institute for Quantum Computing and Dept. of Combinatorics \& Optimization \\ University of Waterloo and St. Jerome's University, \\ and Perimeter Institute for Theoretical Physics \\
www.iqc.ca/ $\tilde{\hspace{2mm} }$ mmosca/web}
\vspace{5mm}

\end{center}

%\end{abstract}
%\maketitle
\sloppy

\section*{Article Outline}

Glossary
\begin{enumerate}

\item Definition of the Subject and Its Importance

\item Introduction and Overview

\item The Early Quantum Algorithms

\item Factoring, Discrete Logarithms, and the Abelian Hidden Subgroup Problem

\item Algorithms based on Amplitude Amplification

\item Simulation of Quantum Mechanical Systems

\item Generalizations of the Abelian Hidden Subgroup Problem

\item  Quantum Walk Algorithms

\item  Adiabatic Algorithms

\item  Topological Algorithms

\item Quantum algorithms for quantum tasks

\item Future Directions

\item Bibliography

    \end{enumerate}

\section*{Glossary}

{\bf Quantum Circuit Model:} One of the standard and most commonly used models of quantum computation which generalizes the classical model of acyclic circuits and closely models most of the proposed physical implementations of quantum computers.
When studying algorithms for a problem with an infinite number of possible inputs, one usually restricts attention to {\it uniform} families of circuits, which are families of circuits in which the circuit for inputs of size $n$ can be generated efficiently as a function of $n$. For example, one might require that there is a classical Turing machine that can generate the $n$th circuit in time polynomial in $n$.

{\bf Black Box Model:} A model of computation where the input to the problem includes a ``black-box'' that can be applied (equivalently, an ``oracle'' that can be ``queried''). This is the only way to extract information from the black-box. For example, the black-box could accept inputs $j \in \{0,1\}^n$ and output a value $X_j \in \{0,1\}$. In this particular case, we can think of the black-box as a means for querying the bits of the string $\X = X_1 X_2 X_3 \ldots X_{2^n}$. In the black-box model, one usually measures complexity in terms of the number of applications of the black-box.

{\bf Computational Complexity:} When referring to an algorithm, the computational complexity (often just called the complexity) is a measure of the resources used by the algorithm (which we can also refer to as the {\it cost} of the algorithm) usually measured as a function of the size of the input to the algorithm. The complexity for input size $n$ is taken to be the cost of the algorithm on a worst-case input to the problem of size $n$. This is also referred to as {\it worst-case complexity}. When referring to a problem, the computational complexity is the minimum amount of resources required by any algorithm to solve the problem. See \cite{Wat08} for an overview.

{\bf Query Complexity:} When referring to a black-box algorithm, the query complexity is the number of applications of the black-box or oracle used by the algorithm. When referring to a black-box problem, the query complexity is the minimum number of applications of the black-box required by any algorithm to solve the problem.

\section{Definition of the Subject and Its Importance}

The Strong Church-Turing thesis states that a probabilistic Turing machine can efficiently simulate any realistic model of computation. By ``efficiently'', we mean that there is a polynomial $p$ such that the amount of resources used by the Turing machine simulation is not more than $p(M)$ where $M$ is the amount of resources used by the given realistic model of computation.

Since a computer is a physical device, any reasonable model of computation must be cast in a realistic physical framework, hence the condition that the model be ``realistic'' is very natural. The probabilistic Turing machine is implicitly cast in a classical framework for physics, and it appears to hold as long as the competing model of computation is also cast in a classical framework. However, roughly a century ago, a new framework for physics was developed, quantum mechanics. The impact of this new framework on the theory of computation was not take very seriously until the early 1970's by Stephen Wiesner \cite{Wie71} for cryptographic purposes (and later by Bennett and Brassard \cite{BB84}). Benioff \cite{Ben82} proposed using quantum devices in order to implement reversible computation. Feynman \cite{Fey82} noted that a classical computer seems incapable of efficiently simulating the dynamics of rather simple quantum mechanical systems, and proposed that a ``quantum'' computer, with components evolving according to quantum mechanical rules, should be able to perform such simulations efficiently (see section \ref{simulation}). Manin made similar observation independently \cite{Man00}. Deutsch \cite{Deu85} worked on proving the original Church-Turing thesis (which was only concerned about effective computability, and not efficient computability) in a quantum mechanical framework, and defined two models of quantum computation; he also gave the first quantum algorithm. One of Deutsch's ideas is that quantum computers could take advantage of the computational power present in many ``parallel universes'' and thus outperform conventional classical algorithms. While thinking of parallel universes is sometimes a convenient way for researchers to invent quantum algorithms, the algorithms and their successful implementation are independent of any particular interpretation of standard quantum mechanics.

Quantum algorithms are algorithms that run on any realistic model of quantum computation. The most commonly used model of quantum computation is the circuit model (more strictly, the model of uniform families of acyclic quantum circuits), and the {\it quantum Strong Church-Turing thesis} states that the quantum circuit model can efficiently simulate any realistic model of computation. Several other models of quantum computation have been developed,
 and indeed they can be efficiently simulated by quantum circuits. Quantum circuits closely resemble most of the currently pursued approaches for attempting to construct scalable quantum computers.

The study of quantum algorithms is very important for several reasons.
Computationally secure cryptography is widely used in society today, and relies on the believed difficulty of a small number of computational problems.  Quantum computation appears to redefine what is a tractable or intractable problem, and one of the first breakthroughs in the development of quantum algorithms was Shor's discovery of efficient algorithms \cite{Sho94} for factoring and finding discrete logarithms.  The difficulty of factoring and finding discrete logarithms was (and still is!) at the core of currently-used public-key cryptography, and his results showed that if and when a quantum computer is built, then any messages that had been previously encrypted using our current public-key cryptographic tools could be compromised by anyone who had recorded the ciphertext and public keys. Furthermore the vast public-key infrastructure currently in place would be compromised with no clear alternative to replace it; also, any alternative will take many years to deploy.
There was a sudden rush to answer two fundamental questions. Firstly, can we actually build a sufficiently large quantum computer? Perhaps this isn't a reasonable model of computation. Subsequent work on quantum fault-tolerant error correction indicates that the answer is ``yes'', and experimental progress has steadily grown. Secondly, what other interesting and important problems can quantum computers solve more efficiently than the best known classical algorithms? The following sections survey the state of the art on the second question.

As we gain confidence about which problems are still hard in a quantum mechanical framework, we can start to rebuild confidence in a secure public-key cryptographic infrastructure that is robust in the presence of quantum technologies.
Although the cryptographic implications are dramatic and of immediate relevance, in the longer term, the most important significance of quantum algorithms will be for a wider range of applications, where important problems cannot be solved because there are no known (or possible) efficient classical algorithms, but there are efficient quantum mechanical solutions. At present, we know of applications such as searching and optimizing (section \ref{search}) and simulating physical systems (see section \ref{simulation}), but the full implications are still unknown and very hard to predict. The next few sections give an overview of the current state of quantum algorithmics.

\section{Introduction and Overview}

There are several natural models of quantum computation.
The most common one is a generalization of the classical circuit model. A detailed description of the circuit model of quantum computation can be found in several textbooks \cite{NC00, KSV02, KLM06}.
One can also define continuous models of computation, where one specifies the Hamiltonian $H(t)$ of the system at time $t$, where $H(t)$ is a ``reasonable'' Hamiltonian (e.g. a sum Hamiltonians involving a constant number of nearby subsystems), and one evolves the system for a period of time $T$. A reasonable measure of the total cost might be $\int_{t=0}^T  || H(t)|| dt $.

Most algorithmic work in quantum computing has been developed in discrete models of computation, that is, with a discrete state space and with discrete time steps.
In sections \ref{continuous-walks} and \ref{adiabatic}, we discuss algorithms developed in a continuous-time model of computation. Even if, as in the case of classical computers, implementations of scalable fault-tolerant quantum computers have discrete time steps and state spaces, these algorithms are still very useful since there are efficient simulations using any universal discrete model of computation. Note that if no such efficient simulation exists, then either the continuous model of computation in question is physically unrealistic, or the quantum Strong Church-Turing thesis is incorrect.

Discrete state spaces can be used to approximate continuous state spaces in order to solve problems normally posed with continuous state spaces. One must choose appropriate discretizations, analyzing errors in the approximation, and quantify the scaling of the algorithm as the overall approximation error gets arbitrarily small. Quantum algorithms for such continuous problems are surveyed in \cite{traub07}.

Many of the key ideas that led to the development of quantum computation emerged from earlier work on reversible computation \cite{Ben88}. Many facts from the theory of reversible computing are fundamental tools in the development of quantum algorithms.
For example, suppose we have a classical algorithm for computing a function $f : \{0,1\}^n \rightarrow \{0,1\}^m$ (we use binary encoding for convenience). The details of the computing model are not so important, as long as it's a realistic model. For concreteness, suppose we have a reasonable encoding of a circuit of size $C$, using gates from a finite set, that takes $\x \in \{0,1\}^n$ as input and outputs $\y \in \{0,1\}^m$ (discarding any additional information it might have computed).  Then this circuit can be efficiently converted into a circuit, composed only of reversible gates, that maps $\ket{\x} \ket{\y} \ket{\0} \mapsto \ket{\x} \ket{\y \oplus f(\x)} \ket{\0}$, where $\oplus$ denotes the bitwise XOR (addition modulo $2$), and the third register of $0$s is ancilla workspace that is reset to all $0$s by reversible operations. This new circuit uses $O(C)$ reversible gates from a finite set, so the overhead is modest. A basic introduction to this and other important facts about reversible computing can be find in most standard textbooks on quantum computing. For example, in section \ref{search} on quantum searching, we use the fact that we can convert any classical heuristic algorithm that successfully guesses a solution with probability $p$ into a reversible quantum algorithm that guesses a solution with probability amplitude $\sqrt{p}$.

Most of the known quantum algorithms can be phrased as black-box algorithms solving black-box problems.
A black-box, or oracle, is subroutine or subcircuit that implements some operation or function. It does so in a way that provides no other information other than simply taking an input and giving the prescribed output. One cannot, for example, look at the inner workings of the circuit or device implementing the black-box to extract additional information.
 For example, Shor's factoring algorithm can be viewed as an algorithm that finds the order of an element in a black-box group (that is, a group for which the group operations are computed by a black-box), or the period of a black-box function, where the black-box is substituted with a subcircuit that implements exponentiation modulo $N$. The quantum search algorithm is described as a black-box algorithm, but it is straightforward to substitute in a subcircuit that checks if a given input is a valid certificate for some problem in NP.

 If we take a black-box algorithm that uses $T$ applications of the black-box and $A$ other computational steps, and replace each black-box with a subcircuit that uses $B$ elementary gates, then we get an algorithm that uses $TB+A$ gates.
 Thus if $T$ and $A$ are both polynomial in size, then an efficient black-box algorithm yields an efficient algorithm whenever we replace the black-box with a polynomial time computable function.

Many lower bounds have been found in the black-box model. The query complexity of a black-box problem is the number of applications of the black-box (or queries to the oracle) that a black-box algorithm must make in order to solve the problem. If we try to solve a problem that has query complexity $T$, where the black-box is implemented by some subcircuit, then we can conclude that any algorithm that treats the subcircuit as a black-box must apply the subcircuit a total of $T$ times and thus use $\Omega(T)$ gates (we use the fact that any implementation of the black-box uses at least $1$ gate; if we had a lower bound on the complexity of implementing the black-box, then we could derive a better lower bound in this case). However, this does not imply that an $\Omega(T)$ lower bound applies to any algorithm that uses the subcircuit, since it might exploit the information within the subcircuit in a way other than just applying the subcircuit.
A discussion of the black-box model and its practical relevance can be found in \cite{Cle00, KLM06, Vaz98}.

In the literature, one can often seen a progression from studying basic algorithmic primitives (such as the convergence properties of a generic quantum walk), to the application to solve a black-box problem (such as element distinctness), to solving some concrete computational problem (like factoring an integer) that doesn't involve a black-box.

This survey will include both black-box and non-black-box results. It is infeasible to detail all the known quantum algorithms, so a representative sample is given in this article. For a subset of this sample, there is an explicit definition of the problem, a discussion of what the best known quantum algorithm can do, and a comparison to what can be achieved with classical algorithms. For black-box problems, we attempt to give the number of queries and the number of non-query operations used by the algorithm, as well as the best-known lower bounds on the query complexity. In some cases, all of this information is not readily available in the literature, so there will be some gaps.

As a small technical note, when we refer to a real number $r$ as an input or output to a problem, we are referring to a finite description of a real number from which, for any integer $n$, one can efficiently compute (in time polynomial in $n$) an approximation of $r$ with error at most $1/2^n$.
%Also, an $n$-tuple of values $(x_1, x_2, \ldots, x_n)$ will be denoted in bold $\x$

In this article, we start with a brief sketch of the very early quantum algorithms, and then in the subsequent sections the algorithms are grouped according to the kind of problems they solve, or the techniques or algorithmic paradigms used.

Section \ref{early} summarizes the early quantum algorithms.
Section \ref{ahsp} describes the Abelian Hidden Subgroup algorithms, including Shor's factoring and discrete logarithm algorithms.
Section \ref{search} describes quantum searching and amplitude amplification and some of the main applications.
Section \ref{simulation} describes quantum algorithms for simulating quantum mechanical systems, which are another important class of algorithms that appear to offer an exponential speed-up over classical algorithms.
Section \ref{hsp} describes several non-trivial generalizations of the Abelian Hidden Subgroup Problem, and related techniques.
Section \ref{walks} describes the quantum walk paradigm for quantum algorithms and summarizes some of the most interesting results and applications.
Section \ref{adiabatic} describes the paradigm of adiabatic algorithms.
Section \ref{topological} describes a family of ``topological'' algorithms.
Section \ref{quantumtasks} describes algorithms for quantum tasks which cannot be done by a classical computer.
In section \ref{future} we conclude with a discussion.

\section{Early Quantum Algorithms} \label{early}

The first explicitly defined quantum algorithm was the one described by David Deutsch in his landmark paper \cite{Deu85} where he defined the model of quantum computation, including a circuit and Turing machine model.

The problem was to decide if a given function $f : \{0,1\} \mapsto \{0,1\}$ is constant or ``balanced'' (the function $f$ is balanced if there are an equal number of $0$ and $1$ outputs). In other words, output $f(0) \oplus f(1)$, where $\oplus$ denotes addition modulo $2$. One is given a circuit that implements $\ket{x}\ket{0} \mapsto \ket{x} \ket{f(x)}$. Deutsch showed that using one application of the circuit, we can compute \[ \frac{1}{\sqrt{2}} \ket{0} \ket{f(0)} + \frac{1}{\sqrt{2}} \ket{1} \ket{f(1)} .\]
If $f(0) = f(1)$, then the first qubit is in the state $\frac{1}{\sqrt{2}} \ket{0} + \frac{1}{\sqrt{2}} \ket{1}$ and thus  applying a Hadamard transformation to the first qubit will output $\ket{0}$ with certainty. However, if $f(0) \neq f(1)$, then applying a Hadamard gate and measuring will output $\ket{0}$ with probability $\frac{1}{2}$ and $\ket{1}$ with probability $\frac{1}{2}$. Thus if we obtain $\ket{1}$, we are certain that the function is balanced, and thus we would output ``balanced'' in this case. If we obtain $\ket{0}$, we cannot be certain; however, by guessing ``constant'' with probability $\frac{2}{3}$ in this case, we get an algorithm that guesses correctly with probability $\frac{2}{3}$. This is not possible with a classical algorithm that only evaluates $f$ once.

If one knows, for example, that the unitary transformation also maps $\ket{x} \ket{1} \mapsto \ket{x} \ket{1 \oplus f(x)}$, then one can solve this problem with certainty with only one query \cite{CEMM98}.  A common tool that is used in this and many other quantum algorithms, is to note that if one applies such an implementation of $f$ on the input $\ket{x} \left( \frac{1}{\sqrt{2}} \ket{0} - \frac{1}{\sqrt{2}} \ket{1} \right)$, then the output can be written as $(-1)^{f(x)} \ket{x} \left( \frac{1}{\sqrt{2}} \ket{0} - \frac{1}{\sqrt{2}} \ket{1} \right)$.  This technique can be used to replace the $2$-step process of first mapping $\ket{x} \ket{0} \mapsto \ket{x} \ket{f(x)}$, then applying a $Z$ gate $\ket{x} \ket{f(x)} \mapsto \ket{x} (-1)^{f(x)} \ket{f(x)} = (-1)^{f(x)} \ket{x} \ket{f(x)}$, and then applying the function evaluation a second time in order to ``uncompute'' the value of $f(x)$ from the second register: $(-1)^{f(x)} \ket{x} \ket{f(x)} \mapsto (-1)^{f(x)} \ket{x} \ket{0}$.

This problem was generalized to the Deutsch-Jozsa problem \cite{DJ92} which asks the same constant versus ``balanced'' question, but for a function $f: \{0,1\}^n \rightarrow \{0,1\}$, with the promise that the function is either constant or balanced. The notion of a promise problem appears frequently in the study of quantum algorithms and complexity. One can either think of the condition as a promise, or alternatively, one can accept any answer as correct in the case that the promise is not satisfied.

This problem was solved with $2$ queries with a similar algorithm, and the $2$ queries can be reduced to one $1$ if the oracle evaluating $f$ has the form $\ket{x} \ket{b} \mapsto \ket{x} \ket{b \oplus f(x)}$. A classical deterministic algorithm requires $2^{n-1} + 1$ evaluations of the function in order to decide if $f$ is constant or balanced with certainty. However, a classical randomized algorithm can decide if $f$ is constant or balanced with error probability $\epsilon$ with only $O(\log \frac{1}{\epsilon})$ queries.  Other oracle separations were given in \cite{BB92, BB94}.

Bernstein and Vazirani \cite{BV97} defined a special family of functions that are either constant or balanced, in particular, for any $\mathbf{a} \in \{0,1\}^n$, $b \in \{0,1\}$, let $f_{\mathbf{a}} (\x) = \mathbf{a} \cdot \x \oplus b$. They showed how to find $\mathbf{a}$ using only $2$ evaluations of a black-box implementing $\ket{\x} \ket{0} \mapsto \ket{\x} \ket{f(\x)}$. This can be reduced to $1$ evaluation given the standard black-box $U_f : \ket{\x} \ket{b} \mapsto \ket{\x} \ket{b \oplus f(\x)}$. This algorithm can be generalized quite naturally to finding the hidden matrix $M$ in the affine map $\x \mapsto M \x + \mathbf{b}$, where $\x \in \Z_N^n$, $\mathbf{b} \in \Z_N^m$ and $M$ is an $m \times n$ matrix of values in $\Z_N$ (with element-wise addition modulo $N$), using only $m$ queries \cite{Hoy99, Mos99}. In some sense, the quantum algorithm takes a black-box for right-multiplying by $M$ and turns it into a black-box for left-multiplying by $M$. This allows us to determine $M$ with only $m$ queries instead of $n+1$ queries, which can be advantageous if $m < n$, although no practical application has been developed to date.

Bernstein and Vazirani also give the first instance of a black-box problem (recursive Fourier sampling) where a quantum algorithm  gives an exponential improvement over bounded error randomized classical algorithms.
Simon \cite{Sim94} gave another problem where there is an exponential gap. The problem is somewhat simpler to state, and is a special case of a broad family of important problems for which efficient quantum algorithms were subsequently found (using similar methods).

\begin{fminipage}{\mylength}
\textbf{\underline{Simon's problem}} \vskip 6pt

\textbf{Input: } A black-box $U_f$ implementing $\ket{\x}\ket{0} \mapsto \ket{\x} \ket{f(\x)}$ for a function $f:\{0,1\}^n \rightarrow \{0,1\}^n$ with the property that $f(\x) = f(\y)$ if and only if $\x \oplus \y \in K = \{ \0, \s\}$ for some $\s \in \{0,1\}^n$.

\textbf{Problem: } Find $\s$.
\end{fminipage}
Note that $f$ is one-to-one if $\s = \0$, the string of $0$s, and $f$ is two-to-one if $\s \neq \0$.

Simon's algorithm is very simple and elegant. We sketch it here, since many of the algorithms presented in the upcoming sections follow the same overall structure. We give a modified version that has a definite running time (versus letting it run indefinitely and analyzing the expected running time).
\begin{fminipage}{\mylength}
\textbf{\underline{Simon's algorithm}} \vskip 6pt

\begin{enumerate}
\item
Start with the $2n$ qubit state $\ket{00 \ldots 0} \ket{00 \ldots 0}$. Set $i = 1$.

\item
Apply a Hadamard gate to the first $n$ qubits to obtain the state $\frac{1}{\sqrt{2^n}} \ket{\x} \ket{\0}$.

\item \label{state-creation}
Apply $U_f$ to create the state  $\frac{1}{\sqrt{2^n}} \ket{\x} \ket{f(\x)} $.

\item
Apply a Hadamard gate to the first $n$ qubits.

\item
Measure the first $n$ qubits to obtain a string $\y_i \in \{0,1\}^n$.

\item
 If $i = n+3$, go to the next step. Otherwise, increment $i$ and go to step 2.

\item
Let $M$ be the $(n+3)\times n$ matrix whose $i$th row is the vector $\y_i$.
Solve the system $M \x = \0$ (where we treat $\x$ and $\0$ as column vectors).  If there is a unique non-zero solution $\x = \s$, then output $\s$. If the only solution is $\x = \0$, then  output $\0$.  Otherwise, output ``FAIL''.

\end{enumerate}
\end{fminipage}

Note that in step \ref{state-creation}, since we can partition the space $\Z_2^n$ into a disjoint union of cosets $\Z_2^n/K$ of the subgroup $K = \{ \0, \s \}$, where $f(\x)$ is constant on each coset, we can rewrite the state of the system as
\[ \frac{1}{\sqrt{2^n}} \ket{\x} \ket{f(\x)} = \frac{1}{\sqrt{2^{n-1}}} \sum_{ \{\z, \z \oplus \s \}  \in \Z_2^n/K} \frac{1}{\sqrt{2}} (\ket{\z} + \ket{\z \oplus \s}) \ket{f(\z)} .\]

Since the rest of the algorithm ignores the second register, we can assume that in the first register we have a random ``coset state'' $\frac{1}{\sqrt{2}} (\ket{\z} + \ket{\z \oplus \s} )$ selected uniformly at random from all the coset states.

After applying the Hadamard gates, the coset state gets mapped to a equally weighted superposition of states that are orthogonal to $K$, \[ \frac{1}{\sqrt{2^{n-1}}} \sum_{\y \in K^{\perp}} (-1)^{\y \cdot \z} \ket{\y} \] where $K^{\perp} = \{ \y \mid \y \in \Z_2^n, \y \cdot \s = 0 \}$. Thus, with $n + O(1)$ random samples from $K^{\perp}$, the samples vectors $\y_i$ will generated $K^{\perp}$ with high probability. Then, using linear algebra, we can efficiently compute generators for $K$.
In the generalizations of Simon's algorithm that we see in the next few sections, we'll see a more general formulation of $K^{\perp}$ for a hidden subgroup $K$.

Shortly after Simon came up with his black-box algorithm, Shor \cite{Sho94} used similar ideas to derive his famous algorithms for factoring and finding discrete logarithms.
\begin{fminipage}{\mylength}
\textbf{\underline{Quantum algorithms for Simon's problem}} \vskip 6pt
Simon's algorithm solves this problem with bounded error using $n + O(1)$ applications of $U_f$ and $O(n)$ other elementary quantum operations and $O(n^3)$ elementary classical operations.

Brassard and H{\o}yer \cite{BH97} combined Simon's algorithm with amplitude amplification order to make the algorithm exact.
\end{fminipage}

\begin{fminipage}{\mylength}
\textbf{\underline{Classical algorithms for Simon's problem}} \vskip 6pt
Simon \cite{Sim94} showed a lower bound of $\Omega(2^{\frac{n}{4}})$ queries, and this can be improved to $\Omega(2^{\frac{n}{2}})$.
\end{fminipage}

This was a historical moment, since the field of quantum algorithms progressed from work on black-box algorithms and complexity, to an algorithm without black-boxes and of broad practical importance. It is important to note the fundamental role of the foundational complexity-theoretic work and black-box algorithms to the development of the practical algorithms.

\section{Factoring, Discrete Logarithms and the Abelian Hidden Subgroup Problem} \label{ahsp}

\subsection{Factoring, Finding Orders and Periods, and Eigenvalue Estimation}

The most well known quantum algorithm is Shor's algorithm \cite{Sho94, Sho97} for the integer factorization problem.

\begin{fminipage}{\mylength}
\textbf{\underline{Integer factorization problem}} \vskip 6pt

\textbf{Input: } An integer $N$.

\textbf{Problem: } Output positive integers $p_1, p_2, \ldots, p_l,
r_1, r_2, \ldots, r_l$ where the $p_i$ are distinct primes and $N =
p_1^{r_1} p_2^{r_2} \ldots p_l^{r_l}$.
\end{fminipage}

This problem can be efficiently reduced (that is, in time polynomial in $\log N$), by a probabilistic classical algorithm, to $O(l)$ instances (note that $l \in O(\log N)$) of the problem of finding the multiplicative order of an element modulo $N$, which can be solved efficiently on a quantum computer.

\begin{fminipage}{\mylength}
\textbf{\underline{Order finding problem}} \vskip 6pt

\textbf{Input: } Positive integers $a$ and $N$, such that
$\GCD(a,N)=1$ (i.e. $a$ is relatively prime to $N$).

\textbf{Problem: } Find the order of $a$ modulo $N$.
\end{fminipage}

Essentially the same quantum algorithm can efficiently find the order of an element $a$ in a finite group $G$ given a black-box for performing the group arithmetic.  Shor's algorithm in fact solves the more general problem of finding the period of a periodic function $f$.

\begin{fminipage}{\mylength}
\textbf{\underline{Period finding problem}} \vskip 6pt

\textbf{Input: } A black-box implementing a periodic function $f:\mathbb{Z} \mapsto X$ for some finite set $X$, where $f(x) = f(y)$ if and only if $r \mid x-y$.

\textbf{Problem: } Find the period $r$.
\end{fminipage}

Shor described his algorithm for the specific function $f(x) = a^x \mod N$, where $N$ was the integer to be factored, however the algorithm will find the period of any such periodic function (note the assumption that the values of $f(1), f(2), \ldots, f(r)$ are distinct; one can also analyze the case of where the values are not entirely distinct \cite{BL95, ME98}).

 Kitaev later showed that the problem of finding the order of $a \in G$ can alternatively be reduced to the problem of estimating the eigenvalues of the operator that multiplies by $a$, and he described a efficient quantum algorithm for estimating such eigenvalues \cite{Kit95}. His method is to reduce the problem to phase estimation. Although Kitaev's algorithm was qualitatively different \cite{Joz98}, it was shown that using an improved phase estimation algorithm yields an order-finding circuit that is essentially equivalent to Shor's factoring algorithm based on period-finding \cite{CEMM98}.

Both Shor's and Kitaev's approaches find $r$ by finding good estimates of a random integer multiple of $\frac{1}{r}$ and then applying the continued fractions algorithm to find $r$.

\begin{fminipage}{\mylength}
\textbf{\underline{Sampling estimates}}\newline \textbf{\underline{
to an almost uniformly random integer multiple of $\frac{1}{r}$}}
\vskip 6pt

\textbf{Input: } Integers $a$, and $N$ such that $\GCD(a,N) =1$. Let
$r$ denote the (unknown) order of $a$.

\textbf{Problem: }    Output a number $x \in \{0,1,2, \ldots,
2^{n}-1 \}$ such that for each $k\in\{0,1,\ldots, r-1\}$ we have
\[\Pr\left(\left|
\frac{x}{2^n} - \frac{k}{r} \right| \leq \frac{1}{2r^2}\right)\geq
c\frac{1}{r}\] for some constant $c > 0$.
\end{fminipage}

Shor's analysis of the algorithm works by creating a ``periodic state'', which is done by preparing $\sum_x \ket{x}\ket{a^x \mod N}$ and measuring the second register (in fact, just ignoring or tracing out the second register suffices, since one never uses the value of the measurement outcome). One then applies the quantum Fourier transform, or its inverse, to estimate the period.

Kitaev's algorithm works by estimating a random eigenvalue of the operator $U_a$ that multiplies by $a$.

\begin{fminipage}{\mylength}
\textbf{\underline{Eigenvalue estimation problem}} \vskip 6pt

\textbf{Input:} A quantum circuit implementing the controlled-$U$, for some unitary operator $U$, and
an eigenstate $\ket{\psi}$ of $U$ with eigenvalue $e^{2\pi i
\omega}$.

\textbf{Problem:} Obtain a good estimate for $\omega$.
\end{fminipage}

He solves the eigenvalue estimation problem by solving a version of the well-studied phase estimation problem, and uses the crucial fact (as did Shor) that one can efficiently implement a circuit that computes $U_a^{2^j} = U_{a^{2^j}}$ using $j$ group multiplications instead of $2^j$ multiplications. Thus one can efficiently reduce the eigenvalue estimation problem to the following phase estimation problem.

\begin{fminipage}{\mylength}
\textbf{\underline{Phase estimation problem}} \vskip 6pt

\textbf{Input:} The states $\frac{1}{\sqrt{2}} (\ket{0} + e^{2 \pi i \omega y} \ket{1})$, for $y = 1,2, 4, \ldots, 2^n$, for some $\omega \in [0, 2\pi)$.

\textbf{Problem:} Obtain a good estimate of the phase parameter
$\omega$.
\end{fminipage}

Kitaev's phase estimation algorithm used $O(n)$ copies of the states $\frac{1}{\sqrt{2}} (\ket{0} + e^{2 \pi i \omega y} \ket{1})$ for $y = 1, 8, 64, \ldots$, and provides an estimate with error at most $\frac{1}{2^n}$ with high probability. Although there are slightly more efficient phase estimation algorithms, one advantage of his algorithm is that it does not require a quantum Fourier transform, and instead performs some efficient classical post-processing of estimates of $y \omega \mod 1$. This might be advantageous experimentally.
The phase estimation problem has been studied for several decades \cite{helstrom, holevo} recently with some focus on the algorithmic complexity of the optimal or near-optimal estimation procedures.

\begin{fminipage}{\mylength}
\textbf{\underline{Quantum algorithms for order finding}} \vskip 6pt

Finding the order of a random element in $Z_N^*$
\begin{itemize}
\item[-] Quantum complexity is in $O((\log N)^2 \log\log(N) \log\log\log(N)
)$.
\end{itemize}

Order finding in a black-box group
\begin{itemize}
\item[-]
Quantum black-box complexity (for groups with unique encodings
of group elements) is $O(\log r)$ black-box multiplications and
$O(n+\log^2 r)$ other elementary operations.
\end{itemize}

\end{fminipage}

\begin{fminipage}{\mylength}
\textbf{\underline{Classical algorithms for order finding}} \vskip 6pt

Finding the order of a random element in $Z_N^*$
\begin{itemize}
\item[-] Best known rigorous probabilistic classical algorithm has
complexity in $e^{O(\sqrt{\log N \log \log N})}$.
\item[-] Best known heuristic probabilistic classical algorithm has
complexity in $e^{O( (\log N)^{\frac{1}{3}} (\log\log
N)^{\frac{2}{3}})}$.
\end{itemize}

 Order finding in a black-box group
\begin{itemize}
\item[-] Classical black-box multiplication complexity is in $\Theta(\sqrt{r})$. \cite{Cle99}
\end{itemize}
\end{fminipage}
By ``heuristic'' algorithm, we mean the proof of its
running time makes some plausible but unproven assumptions.

\subsection{Discrete Logarithms} \label{DLP-section}

Shor \cite{Sho94,Sho97}  also solved the problem of finding discrete logarithms in the multiplicative group of a finite field.

\begin{fminipage}{\mylength}
\textbf{\underline{The discrete logarithm problem}} \vskip 6pt

\textbf{Input:} Elements $a$ and $b=a^t$ in $\mathbb{Z}_p^{*}$,
where $t$ is an integer from $\{0,1,2, \ldots, r-1\}$ and $r$ is the
order of $a$.

\textbf{Problem:} Find $t$.  (The number $t$ is called the discrete
logarithm of $b$ with respect to the base $a$.)
\end{fminipage}

Shor solved this problem by defining $f: \mathbb{Z}_r \times \mathbb{Z}_r \mapsto \mathbb{Z}_p^{*}$ as $f(x,y) = a^x b^y$. Note that $f(x_1, y_1) = f(x_2, y_2)$ if and only if $(x_1 - x_2, y_1 - y_2)$ is in the additive subgroup of $\mathbb{Z}_r \times \mathbb{Z}_r$ generated by $(t, -1)$. Since $r$ is known, there is no need for the continued fractions algorithm. There is also an analogous eigenvalue estimation version of this algorithm.

The algorithm can be defined in general for any group $G$ where we have a black-box for computing the group operation. Given $a , b\in G$, output the smallest positive $t$ such that $b = a^t$.  For example, one can apply this algorithm to finding discrete logarithms in the additive group of an elliptic curve over a finite field \cite{PZ03, Kay05}, a group widely used in public key cryptography \cite{MVV96}. In this case, the group operation is described as addition, and thus the problem is to find the smallest positive integer $t$ such that $b = ta$, where $b$ and $a$ are points on some elliptic curve.

\begin{fminipage}{\mylength}
\textbf{\underline{Quantum complexities of the discrete logarithm problem}}
\vskip 6pt
\begin{itemize}
\item
Finding discrete logarithms in $F_q^*$
\begin{itemize}
\item[-] Quantum complexity is in $O((\log q)^2 \log\log(q) \log\log\log(q)
)$.
\end{itemize}
\item Discrete logarithms in a black-box group represented with
strings of length $n$ (including elliptic curve groups discussed above)
\begin{itemize}
\item[-] Quantum black-box complexity (for groups with unique encodings of group elements)
is $O(\log r)$ black-box multiplications and $O(n+\log^2 r)$ other
elementary operations.
\end{itemize}
\end{itemize}
\end{fminipage}

\begin{fminipage}{\mylength}
\textbf{\underline{Classical complexities of the discrete logarithm problem}}
\vskip 6pt
\begin{itemize}
\item
Finding discrete logarithms in $\mathbf{F}_q^*$
\begin{itemize}
\item[-] Best known rigorous probabilistic classical algorithm has
complexity in $e^{O(\sqrt{\log q \log \log q})}$ for certain values of $q$ (including $q=2^n$ and prime $q$).
\item[-] Best known heuristic
 probabilistic classical algorithm has
complexity in $e^{O( (\log N)^{\frac{1}{3}} (\log\log
N)^{\frac{2}{3}})}$.
\end{itemize}
\item Discrete logarithms in a black-box group represented with
strings of length $n$
\begin{itemize}
\item[-] Classical black-box complexity is in $\Theta(\sqrt{r})$.
For a large class of elliptic curves, the best known classical algorithms have complexity in $O(\sqrt{r})$ group additions. There are sub-exponential algorithms for special families of curves.
\end{itemize}
\end{itemize}
\end{fminipage}

\subsection{Abelian Hidden Subgroup Problem} \label{AHSP}

Notice how we can rephrase most of the problems we have already
discussed, along with some other ones, as a special case of the following problem.

\begin{fminipage}{\mylength}
\textbf{\underline{The Abelian hidden subgroup problem}} \vskip 6pt

Let $f: G \rightarrow X$ map an Abelian group $G$ to some finite set $X$ with
the property that there exists some subgroup $K \leq G$ such that
for any $x,y \in G$, $f(x) = f(y)$ if and only if $x+K = y+K$. In
other words $f$ is constant on cosets of $K$ and distinct on
different cosets.
\end{fminipage}

\begin{enumerate}
\item[]\textbf{Deutsch's Problem:}

$G = \mathbb{Z}_2$, $X= \{0,1\}$, and $K = \{0\}$ if $f$ is
balanced, and $K = \{0,1\}$ if $f$ is constant.

\vskip 6pt
\item[]\textbf{Generalized Simon's problem:}

$G = \mathbb{Z}_2^n$, $X = \{0,1\}^n$, and $K$ is any subgroup of
$\mathbb{Z}_2^n$.

\vskip 6pt
\item[]\textbf{Finding orders:}

$G = \mathbb{Z}$, $X$ is any finite group $H$,  $r$ is the order of
$a \in H$. The subgroup $K = r\mathbb{Z}$ is the hidden subgroup of
$G$, and a generator for $K$ reveals $r$.

\vskip 6pt
\item[]\textbf{Finding the period of a function:}

$G = \mathbb{Z}$, $X$ is any set, $r$ is the period of $f$. The subgroup $K = r\mathbb{Z}$ is
the hidden subgroup of $G$, and a generator for $K$ reveals the
period $r$.

\vskip 6pt
\item[]\textbf{Discrete logarithms in any group:}
$G=\mathbb{Z}_r\times\mathbb{Z}_r$, $X$ is any group $H$.  Let $a$ be
an element of $H$ with $a^r=1$ and suppose $b=a^k$.  Consider the
function $f(x_1,x_2)=a^{x_1}b^{x_2}$.  We have
$f(x_1,x_2)=f(y_1,y_2)$ if and only if
$(x_1,x_2)-(y_1,y_2)\in\{(tk,-t),t=0,1,\ldots,r-1\}$. The hidden
subgroup $K$ is the subgroup generated by $(k,-1)$, where $k$ is the
discrete logarithm.

\vskip 6pt
\item[]\textbf{Hidden linear functions \cite{BL95}:}
$G=\mathbb{Z}\times\mathbb{Z}$.  Let $g$ be some permutation of
$\mathbb{Z}_N$ for some integer $N$. Let $h$ be a function from
$\mathbb{Z}\times\mathbb{Z}$ to $\mathbb{Z}_N$ defined by
$h(x,y)=x+ay\mod N$.  Let $f=g\circ h$.  The subgroup $K$ is the
hidden subgroup generated by $(-a,1)$, and the generator reveals the
hidden linear function $h$.

\vskip 6pt
\item[]\textbf{Self-shift-equivalent polynomials \cite{Gri97}:} Given a
polynomial $P$ in $l$ variables $X_1,X_2,\ldots,X_l$ over
$\mathbb{F}_q$ (the finite field with $q$ elements), the function
$f$ which maps $(a_1,a_2,\ldots,a_l)\in\mathbb{F}_q^l$ to
$P(X_1-a_1,X_2-a_2,\ldots,X_l-a_l)$ is constant on cosets of a
subgroup $K$ of $\mathbb{F}_q^l$.  This subgroup $K$ is the set of
self-shift-equivalences of the polynomial $P$.

\vskip 6pt
\item[]\textbf{Abelian stabilizer problem \cite{Kit95}:}
Let $G$ be any group acting  on a finite set $X$.  That is, each
element of $G$ acts as a map from $X$ to $X$ in such a way that for
any two elements $a,b\in G$, $a(b(x))=(ab)(x)$ for all $x\in X$. For
a particular element $x\in X$, the set of elements which fix $x$
(that is the elements $a\in G$ such that $a(x)=x$) form a subgroup.
This subgroup is called the stabilizer of $x$ in $G$, denoted
$St_G(x)$.  Let $f_x$ denote the function from $G$ to $X$ which maps
$g\in G$ to $g(x)$.  The hidden subgroup of $f_x$ is $St_G(x)$.

\end{enumerate}

If we restrict attention to
finite Abelian groups, or more generally, finitely generated Abelian
groups, then we can efficiently solve the hidden subgroup problem, by generalizations of the algorithms for factoring, finding discrete logarithms, and Simon's problem.

The Abelian Hidden Subgroup problem can also be used to decompose a finite Abelian group into a direct sum of cyclic groups if there is a unique representative for each group element \cite{Mos99, CM01}. For example, the multiplicative group of integers modulo $N$ is an Abelian group, and we can efficiently perform computations in the group. However, having a decomposition of the group would imply an efficient algorithm for factoring $N$.  The class group of a number field is another Abelian group for which a decomposition is believed to be hard to find on a classical computer.
For example, such a decomposition would easily give the size of the class group, which is also known to be as hard as factoring, assuming the Generalized Riemann Hypothesis.
 Computing the class group of a complex number field is a simple consequence of the algorithm for decomposing an Abelian group into a direct sum of cyclic groups, since there are techniques for computing unique representives in this case.
A unique classical representation is sufficient, but a quantum state could also be used to represent a group element. For example, a uniform superposition over classical representives of the same group element would also work (this technique was applied by Watrous in the case of solvable groups \cite{Wat01}).
Computing the class number of a real number field is not as straightforward, since there is no known way to efficiently compute a unique classical representative for each group element. However Hallgren \cite{Hal07} used the techniques outlined in section \ref{lattice} to show how to compute quantum states to represent the group elements, assuming the Generalized Riemann Hypothesis, in the case of class group of a real number field of constant degree, and thus is able to compute the class group in these cases as well .

\begin{fminipage} {\mylength}
\textbf{\underline{Quantum algorithms for the Abelian Hidden Subgroup Problem}} \vskip 6pt
There exists a bounded-error quantum algorithm for finding
generators for the hidden subgroup $K \leq G = Z_{N_1} \times
Z_{N_2} \times \cdots \times Z_{N_l}$ of $f$ using $O(l)$
evaluations of $f$ and $O(\log^3 N)$ other elementary operations, where $N = N_1 N_2 \ldots N_l = |G|$.

It can be shown that $\Omega(l)$ queries are needed for worst-case $K$.
\end{fminipage}

\begin{fminipage} {\mylength}
\textbf{\underline{Classical algorithms for the Abelian Hidden Subgroup Problem}} \vskip 6pt
In the black-box model $\Omega(\sqrt{|G/K|})$ queries are necessary in order to even decide if the hidden subgroup is trivial.
\end{fminipage}

Generalizations of the Abelian Hidden Subgroup Problem are discussed in section \ref{hsp}.

\section{Algorithms based on Amplitude Amplification} \label{search}

In 1996 Grover \cite{Gro96} developed a quantum algorithm for solving the search problem that gives a quadratic speed-up over the best possible classical algorithm.

\begin{fminipage} {\mylength}
\textbf{\underline{Search problem}} \vskip 6pt

\textbf{Input:}
A black-box $U_f$ for computing an unknown function $f:\{0,1\}^n \rightarrow \{0,1\}$.

\textbf{Problem:}
Find a value $\x \in \{0,1\}^n$ satisfying $f(\x)=1$, if one exists. Otherwise, output ``NO SOLUTION''.
\end{fminipage}

The \emph{decision version} of this problem is to output $1$ if there is a solution, and $0$ if there is no solution.

This problem is stated in a very general way, and can be applied to solving a wide range of problems, in particular any problem in $NP$. For example, suppose one wanted to find a solution to an instance $\Phi$ of the $3$-SAT problem. The Boolean formula $\Phi$ is in ``3-conjunctive
normal form'' (3-CNF), which means that it is a
conjunction (logical AND) of clauses, each of which is a disjunction
(logical OR) of three Boolean variables (or their negations). For
example, the following is a 3-CNF formula in the variables $b_1,
b_2, \ldots, b_5$:
\[\Phi=(b_1\vee \bar{b_2}\vee b_5)\wedge(b_1\vee \bar{b_4}\vee
\bar{b_5})\wedge(b_4\vee b_2\vee b_3),\] where we let $\bar{b}$ denote the logical NOT of $b$. A ``satisfying assignment''
of a particular 3-CNF formula $\Phi$ is an assignment of $0$ or $1$
values to each of the $n$ variables such that the formula evaluates
to $1$. Given a satisfying assignment, it is easy to check if it satisfies the formula.
Define $f_{\Phi}$ to be the function that, for any $\x = x_1 x_2 \ldots x_n \in \{0,1\}^n$, maps ${\bf x} \mapsto 1$ if the assignment $b_i = x_i$, $i=1, 2, \ldots, n$ satisfies $\Phi$ and $\x \mapsto 0$ otherwise.  Solving the search problem for $f_{\Phi}$ is equivalent to finding a satisfying assignment for $\Phi$.

If we only learn about $f$ by applying the black-box $U_f$, then any algorithm that finds a solution with high probability for any $f$ (even if we just restrict to functions $f$ with at most one solution) requires $\Omega(\sqrt{2^n})$ applications of $U_f$ \cite{BBBV97}.

The basic intuition as to why a quantum algorithm might provide some speed-up is that a quantum version of an algorithm that guesses the solution with probability $\frac{1}{2^n}$ will in fact produce the correct answer with probability amplitude $\frac{1}{\sqrt{2^n}}$. The hope is that additional guesses increase the amplitude of finding a correction solution by $\Omega(\frac{1}{\sqrt{2^n}})$, and thus $O(\sqrt{2^n})$ guesses and applications of $U_f$ might suffice in order to get the total probability amplitude close to $1$. Of course, one needs to find a unitary algorithm for implementing this, and obvious approaches do not permit the amplitudes to add constructively.
Lov Grover discovered a quantum algorithm that does permit the amplitudes to add up in such a way. This algorithm was analyzed and generalized to the technique known as ``amplitude amplification'' \cite{BBHT98, BH97, BHMT00}. Further generalizations of these search algorithms are discussed in section \ref{walks}.

\begin{fminipage} {\mylength}
\textbf{\underline{Quantum Algorithms for Searching:}} \vskip 6pt
There exists a quantum algorithm that finds a solution with probability at least $\frac{2}{3}$  if one exists, otherwise outputs ``NO SOLUTION'', and uses $O(\sqrt{2^n})$ applications of $U_f$. The algorithm uses $O( n \sqrt{2^n})$ elementary gates. Note that $\Omega(\sqrt{2^n})$ applications of $U_f$ are necessary in order to find a solution for any $f$, even if we restrict attention to functions $f$ with at most one solution.

\vspace{3mm}
If, as an additional input, we are also given an algorithm $A$ that will find a solution to $f(x) = 1$ with probability $p$, then Amplitude Amplification will find a solution with high probability using $O(\frac{1}{\sqrt{p}})$ applications of $U_f$ and of unitary versions of $A$ and $A^{-1}$.

\end{fminipage}

This more general statement implies that if there are $m \geq 1$ solutions, then there is an algorithm which makes an expected number of queries in $O(\sqrt{2^n/m})$ (since guessing uniformly at random will succeed with probability $\frac{m}{2^n}$).
This algorithm works even if the unitary black-box $U_f$ computes $f$ with a small bounded error \cite{HMW03}.

\begin{fminipage} {\mylength}
\textbf{\underline{Classical Algorithms for the Search Problem:}} \vskip 6pt
Any classical algorithm must make $\Omega(2^n)$ queries in order to succeed with probability at least $\frac{2}{3}$ on any input $f$, even if we restrict attention to functions with at most one solution. Exhaustive searching will find a solution using $O(2^n)$ applications of $U_f$.

\vspace{3mm}
If we are also given a black-box $A$ that successfully guesses a solution to $f(x) = 1$ with probability $p \geq \frac{1}{2^n}$, then $\Theta(\frac{1}{p})$ applications of $U_f$ are needed and also sufficient (using random sampling).

\end{fminipage}

If instead of a black-box for $U_f$, we are given a circuit for $f$ or some other description of $f$ (such as a description of the $3$-SAT formula $\Phi$ in the case of $f_{\Phi}$) then there might be more efficient algorithms that uses this additional information to do more than just use it to apply $U_f$. One can directly use amplitude amplification to get a quadratic speed-up for any such heuristic algorithm that guesses a solution with probability $p$, and then repeats until a solution is found.

The quantum searching algorithm has often been called `database' search algorithm, but this term can be misleading in practice. There is a quadratic speed-up if there is an implicit database defined by an efficiently computable function $f$. However, if we are actually interested in searching a physical database, then a more careful analysis of the physical set-up and the resources required is necessary. For example, if a database is arranged in a linear array, then to query an $N$-element database will take time proportional to $N$. If the data is arranged in a square grid, then a general lookup could be possibly be done in time proportional to $\sqrt{N}$.  This limitation is true for both classical and quantum searching, however the hardware assumptions in the two cases are often different, so one needs to be careful when making a comparison.
(e.g. \cite{Zal98}, \cite{RudolphGrover}, and Chapter 6 of \cite{NC00}, discuss these issues.)

Local searching \cite{AA03} restricts attention to architectures or models where subsystems can only interact locally, and does not allow access to any memory location in one unit of time regardless of the memory size (which is not truly possible in any case, but can sometimes be appropriate in practice). The algorithms used include quantum walk algorithms (e.g. \cite{Zha06}) and can achieve quadratic or close to quadratic speed-ups depending on details such as the spatial dimension of the database.

It is also important to note that there has been much work studying the quantum query complexity of searching an \emph{ordered} list. It is known that $\Theta(\log N)$ queries are necessary and sufficient, but the constant factor is not yet known \cite{HNS02, CLP07, BH07}.

\subsection{Other applications of Amplitude Amplification}

Apart from the application to the searching problem, one of the first applications of the quantum searching algorithm was to counting \cite{BHT98}, and more generally amplitude amplification can be used to estimate amplitudes \cite{BHMT00}.
The quantum algorithm for amplitude estimation combines the techniques of the order finding algorithm with amplitude amplification.
Bounds on optimal phase estimation translate to bounds on optimal amplitude estimation. It has several applications, including approximate and exact counting \cite{BHT98, BHMT00}, and approximating the mean (or, in the continuous case \cite{traub07}, the integral) of a function \cite{terhalthesis}. These applications offer up to a quadratic speed-up over classical algorithms.

We have already mentioned the straight-forward, and very useful, application of amplitude amplification to searching: providing a quadratic speed-up up for any algorithm that consists of repeating a subroutine that guesses a solution, until a solution is found.
However, there are many applications of amplitude amplification that are not so straightforward, and require some additional non-trivial algorithmic insights.
Other applications of amplitude amplification include the collision finding problem, which has applications to breaking hash functions in cryptography.

\begin{fminipage} {\mylength}
\textbf{\underline{Collision Finding Problem:}} \vskip 6pt
\textbf{Input:}
A black-box $U_f$ for computing a function $f:\{0,1\}^n \rightarrow \{0,1\}^m$. The function $f$ is $r$ to $1$, for some positive integer $r$.

\textbf{Problem:} Find two distinct $\x, \y \in \{0,1\}^n$ with $f(x) = f(y)$, if $r > 1$. Otherwise output ``NO COLLISION''.
\end{fminipage}

\begin{fminipage} {\mylength}
\textbf{\underline{Quantum Algorithms for Collision Finding}} \vskip 6pt
There exists a quantum algorithm \cite{BHT97} that uses $O(\left(\frac{2^n}{r}\right)^{\frac{1}{3}})$ applications of $U_f$, and $O(\left(\frac{2^n}{r}\right)^{\frac{1}{3}} \polylog\left(\frac{2^n}{r}\right) )$ other elementary operations, and $O(\left(\frac{2^n}{r}\right)^{\frac{1}{3}})$ space, and outputs a collision with probability at least $\frac{2}{3}$ if $r>1$, and outputs ``NO COLLISION'' otherwise.

It can be shown that $\Omega(\left(\frac{2^n}{r}\right)^{\frac{1}{3}})$ applications of $U_f$ are needed.
%MM todo find a reference for the lower bound

\end{fminipage}

\begin{fminipage} {\mylength}
\textbf{\underline{Classical Algorithms for Collision Finding}} \vskip 6pt
There exists a classical probabilistic algorithm that uses $O(\left(\frac{2^n}{r}\right)^{\frac{1}{2}})$ applications of $U_f$, and $O(\left(\frac{2^n}{r}\right)^{\frac{1}{2}}\polylog\left(\frac{2^n}{r}\right))$ other elementary operations, and $O(n)$ space, and outputs a collision with probability at least $\frac{2}{3}$ if $r>1$, and outputs ``NO COLLISION'' otherwise.
It can be shown that $\Omega(\left(\frac{2^n}{r}\right)^{\frac{1}{2}})$ applications of $U_f$ are needed.
\end{fminipage}

Other applications include finding claws \cite{BHT98}, finding the maximum (or minimum) value of a function \cite{hoyerdurr}, string matching \cite{RV01}, estimating the median of a function \cite{Gro98, NW99} and many others \cite{Fur08, DHHM04}.
There is also a non-trivial application to the element distinctness problem \cite{BDHHMSW05}, which we define in section \ref{walks}, since there is a better, optimal, quantum walk algorithm.
Amplitude amplification also a basic tool in making algorithms exact \cite{BH97,MZ04}.

\section{Simulation of quantum mechanical systems} \label{simulation}

Feynman's reason for considering a quantum computer was to simulate other quantum mechanical systems \cite{Fey82}. This general idea was described in more detail by Lloyd \cite{Llo96} and Zalka \cite{Zal98}, and later by many other authors (e.g. \cite{NC00}). More detailed situations have been studied by numerous authors including \cite{SOGKL01, Chi02, BY06, KJLMA08}.

There are other notions of what one what one might mean by simulating a physical system, such as computing properties of the ground state of the Hamiltonian, or other properties of the spectrum of the Hamiltonian.
In this section, we focus on the problem of simulating the evolution of a quantum mechanical system given the initial state (or a description of it) and a description of the Hamiltonian of the system.

For convenience, we will restrict attention to a discrete Hilbert space. In practice, when modelling systems with a continuous state space, the state space will need to be approximated with a discrete state space \cite{Zal98} with enough states so that the accumulated error is below the desired error threshold. This general problem of dealing with continuous problems is discussed in \cite{traub07}.

We will also restrict attention to time-independent Hamiltonians, since well-behaved time-dependent Hamiltonian evolution can be approximated by a sequence of time-dependent Hamiltonian evolutions.

There are two natural ways that have been studied for representing the Hamiltonian.
The first way, is to represent $H = \sum_{k=1}^M H_k$, where $H_k$ is a simple enough Hamiltonian that we know how to efficiently simulate its evolution. For convenience, we assume the simulation of the $H_k$ term for a time $t$ takes unit cost and is exact.
In practice, if the simulation can be done with error $\epsilon$ in time polynomial in $\frac{1}{\epsilon}$, $\tau = ||H_k||t$, and $n$, then the effect on the overall cost is by a factor that is polynomial in these factors. In many situations, the simulation is polynomial in
$\log (\frac{1}{\epsilon})$ and $\log \tau$, and independent of $n$ and $\epsilon$.
 This leads to the following formulation.

\begin{fminipage}{\mylength}
\textbf{\underline{Hamiltonian Simulation Problem 1}} \vskip 6pt
{\bf Input:}

An integer $n$ and black-boxes $A_1$, $A_2, \ldots , A_M$, where $A_j$ takes a non-negative real number $r$ and executes $e^{i H_j r}$, for a set of Hamiltonians $H_j$ acting on $n$ qubits.

The value of $|| H||$, the trace norm of $H = \sum_{k=1}^M H_k$.

A positive real number $t$.

 A positive real number $\epsilon < 1$.

An $n$-qubit state $\ket{\psi}$.

{\bf Output:}

A state $\ket{\psi_f}$ satisfying $|| \ket{\psi_f} - e^{i H t} \ket{\psi} || < \epsilon$.

\end{fminipage}

In practice, the input will likely be a classical description of how to prepare the state $\ket{\psi}$.
It suffices to have an upper bound on $||H||$ (if the bound is within a constant factor, this won't affect the stated complexities).
In particular, it suffices that we can efficiently compute or approximate the eigenvalue $\lambda_k$ of a given an eigenvector $\ket{E_{k}}$ of $H_k$. This is the case, for example, when the Hilbert space is a tensor product of $n$ finite sized subsystems and $H_k$ acts non-trivially only on a finite number of subsystems, say $c$ of them. In other words, up to a reordering of the subsystem labels, $H_k = I^{n-c} \otimes \tilde{H}_k$.
Since
\[ e^{i H_k t} = I^{n-c} \otimes e^{i \tilde{H}_k t} , \] in order to simulate the evolution of $H_k$ for a time interval of size $t$, one only needs to simulate $\tilde{H}_k$ on the relevant $c$ qubits for a time interval of size $t$.
In this case, since we know $\tilde{H}_k$, one can use ``brute-force'' methods to approximately implement the map $\ket{E_{\lambda_k}} \mapsto e^{2 \pi i \lambda_k t}\ket{E_{\lambda_k}}$, for any time interval $t$.

An easy example is when the state space is $n$ qubits, and $H_k$ is a tensor product of Pauli operators. This means that we can easily diagonalize $\tilde{H}_k$ as $\tilde{H}_k = (P_1 \otimes P_2 \otimes \ldots \otimes P_c ) Z^{c} (P_1^{\dagger} \otimes P_2^{\dagger} \otimes \ldots \otimes P_c^{\dagger})$ for some one-qubit unitary operations $P_1, P_2, \ldots, P_c$.
Thus we have
\[ e^{i \tilde{H}_k t} = (P_1 \otimes P_2 \otimes \ldots \otimes P_c) e^{ i Z^{c} t} (P_1^{\dagger} \otimes P_2^{\dagger} \otimes \ldots \otimes P_c^{\dagger}). \]
Since \[ e^{ i Z^{c} t} \ket{x_1 x_2 \ldots x_c} = e^{i t f(x_1 \ldots x_c)} \ket{x_1 x_2 \ldots x_c} \] where $f(x_1, \ldots, x_c) = x_1 \oplus \ldots \oplus x_c$, this simulation can be done easily. In fact, as pointed out in \cite{NC00}, in the case that $H_k$ is such a product Hamiltonian, $c$ does not need to be constant.

Another example, \cite{Zal98}, is where the eigenvectors of $H_k$ are of the form \[ \ket{E_k} = \sum_{j=0}^{2^n-1} e^{2 \pi i jk/2^n} \ket{j} \] (i.e. ``momentum'' eigenstates), in which case the inverse quantum Fourier transform will map $\ket{E_k} \mapsto \ket{k}$, and then one can easily approximate $\ket{k} \mapsto e^{i \lambda_k t} \ket{k}$ for any efficiently computable function $\lambda_k$, and then apply the quantum Fourier transform to map $\ket{k} \mapsto \ket{E_k}$ and thus effectively compute the transformation $\ket{E_k} \mapsto e^{ i \lambda_k t} \ket{E_k}$.

Thus we can study any situation where $H = \sum_k H_k$ and we have some means of efficiently simulating the time evolution of $H_k$. If the $H_k$ commute, then
\[ e^{i H t} = e^{i H_1 t} e^{i H_2 t} \ldots e^{i H_n t} ,  \] and the problem is straightforward. For the non-trivial case when the $H_k$ do not commute pairwise, we can take advantage of approximations derived from Trotter formulas, like
\[ e^{i H t} =  \left(e^{i H_1 t/n} e^{i H_2 t/n} \ldots e^{i H_n t/n} \right)^n + O(t^2/n) \] and other improved versions.  A good estimate of the overall cost of this family of simulations is the number of terms of the form $e^{i H_j r}$ are used, for any choice of $r$ (we can assume $0 \leq r \leq t$, and that $r$ is some efficiently computable number).

\begin{fminipage}{\mylength}
\textbf{\underline{Quantum complexity of Hamiltonian Simulation Problem 1}} \vskip 6pt
There is a quantum algorithm that simulates $e^{i H t}$ on a given input $\ket{\psi}$ with trace distance error at most $\epsilon < 1$ that uses a number of exponential terms $N_{exp}$ satisfying
\[ N_{exp} \in (M \tau + 1)M^{1+o(1)} \tau^{o(1)} \left(\frac{1}{\epsilon}\right)^{O(1/\sqrt{s})}, \]
where $\tau = ||H||t$ and $\epsilon = 1/2^s$.
Since the running time of simulating each exponential term is assumed to be polynomial in $n$, the overall running time is polynomial in $n$.

In general, there are Hamiltonians for which $\Omega(\tau)$ time is necessary.
\end{fminipage}

The product  $\tau =||H|| t$ is the relevant parameter since one can effectively speed up time by a factor of $s$ by rescaling the Hamiltonian by a multiplicative factor of $s$, for any $s > 0$ (e.g. one can perform an operation in half the time by doubling all the energies).

It is hard to give an explicit comparison to the best known classical complexity of this general problem. There are classical algorithms and heuristics for special cases, but in the worst-case, the best solutions known require time exponential in $n$ and polynomial in $\tau$ (e.g. in $\tau^{1 + o(1)}$).

Another natural formulation of the problem \cite{AT07, BACS07} considers the case of sparse Hamiltonians, where for any basis state $\ket{x}$ there are at most $d$ basis states $\ket{y}$ such that $\bra{x}H\ket{y} \neq 0$ and one can efficiently compute this neighbourhood for any input $x$. This leads to the following black-box formulation of the problem.

\begin{fminipage}{\mylength}
\textbf{\underline{Hamiltonian Simulation Problem 2}} \vskip 6pt
{\bf Input:}

Integers $n$ and $d$, and a black-box $U_H$ that maps $\ket{x,i}\ket{0} \mapsto \ket{x,i} \ket{y_i, H_{x,y_i}}$, where $y_i$ is the index of the $i$th non-zero entry in column $x$ of the Hermitian matrix $H$ (if there are $d^{\prime} < d$ non-zero entries, then $U_f$ can output any index for $i > d^{\prime}$). The values $H_{x,y} = \bra{x} H \ket{y}$ are the $(x,y)$ entries of $H$ represented in the computational basis.

A positive real number $t$.

A positive real number $\epsilon < 1$.

An $n$-qubit state $\ket{\psi}$.

{\bf Output:}

A state $\ket{\psi_f}$ satisfying $|| \ket{\psi_f} - e^{i H t} \ket{\psi} || < \epsilon$.
\end{fminipage}

The best known general solution was shown in \cite{BACS07}.

\begin{fminipage}{\mylength}
\textbf{\underline{Quantum complexity of Hamiltonian Simulation Problem 2}} \vskip 6pt
There is a quantum algorithm that simulates $e^{i H t}$ on a given input $\ket{\psi}$ with trace distance error at most $\epsilon < 1$ that uses a number of black-box calls, $N_{bb}$, satisfying, for any positive integer $k$,
\[ N_{bb} \in  O\left(\tau^{(1 + \frac{1}{2k})} d^{(4 + \frac{1}{k})} 5^{2k} \left(\frac{1}{\epsilon}\right)^{\frac{1}{2k}} (\log^* n)\right), \]
where $\tau = ||H||t$ and $\log^* n$ is the smallest positive integer $r$ such that $\log_2^{(r)} n < 2$, where $\log_2^{(r)}$ refers to iterating the $\log_2$ function $r$ times.

In general, there are Hamiltonians that require $\Omega(\tau)$ black-box evaluations.
\end{fminipage}

For example, setting $k = \sqrt{s}$ where $\epsilon = \frac{1}{2^s}$, we get
\[ N_{bb} \in  \tau^{(1 + O(\frac{1}{\sqrt{s}}))} d^{(4 + O(\frac{1}{\sqrt{s}}))} \left(\frac{1}{\epsilon}\right)^{O\left(\frac{1}{
\sqrt{s}}\right)} (\log^* n) . \]

Some applications involve trying to study properties of the ground state of a Hamiltonian by combining simulation of the dynamics with some approach for generating the ground state with non-trivial initial amplitude. For example, if one had a procedure $A$ for generating a ground state $\ket{E_0}$ of $H$ with probability $p$, then one can combine the algorithm for simulating $H$ with eigenvalue estimation, and then amplitude amplification on the eigenvalue estimates in order to generate the ground state with high fidelity. The algorithm would use $O(\frac{1}{\sqrt{p}})$ applications of $A$ and $A^{-1}$ and $O(\frac{1}{\sqrt{p}})$ eigenvalue estimations.
The eigenvalue estimations would have to be precise enough to distinguish $\ket{E_0}$ from the next highest energy eigenstate produced by $A$. If the gap between these two energies is $\Delta$, then the eigenvalue estimation would involve simulating the evolution of $H$ for a time in $\Omega(\frac{1}{\Delta})$.

Several papers address techniques for generating the ground states for various problems of interest including \cite{AL99}.
Another application for such simulations is to implement algorithms that are designed in the continuous-time model in a discrete-time quantum computation model.

\section{Generalizations of the Abelian Hidden Subgroup Problem} \label{hsp}

\subsection{Non-Abelian Hidden Subgroup Problem}

One of the most natural ways to generalize the Abelian Hidden Subgroup Problem is to non-Abelian groups.
The problem definition is the same, apart from letting $G$ be non-Abelian.
One natural problem in this class of problems is the graph automorphism problem.

\vskip 6pt
\textbf{Graph automorphism problem:} Consider $G=S_n$, the
symmetric group on $n$ elements, which corresponds to the
permutations of $\{1,2,\ldots,n\}$.  Let $\G$ be a graph on $n$
vertices labeled $\{1,2,\ldots,n\}$.  For any permutation $\sigma\in
S_n$, let $f_\G$ map $S_n$ to the set of $n$-vertex graphs by
mapping $f_\G(\sigma)=\sigma(\G)$, where $\sigma(\G)$ is the graph
obtained by permuting the vertex labels of $\G$ according to
$\sigma$.  For the function $f_\G$, the hidden subgroup of $G$ is
the automorphism group of $\G$.

The graph isomorphism problem (deciding if two graphs $\G_1$ and $\G_2$ are isomorphic) can be reduced to solving the graph automorphism problem.
The best known classical algorithm takes time in $e^{O(\sqrt{n \log n})}$ to decide if two graphs are isomorphic, and there is no substantially better quantum algorithm known.

There has been much work attacking the non-Abelian hidden subgroup problem.
Ettinger, H{\o}yer and Knill \cite{EHK04} showed the following.
\begin{fminipage}{\mylength}
\textbf{\underline{Query complexity of the non-Abelian Hidden Subgroup Problem}} \vskip 6pt
For any finite group $G$, the non-Abelian hidden subgroup problem can be solved with high probability using $O(\log |G|)$ queries to $U_f$.
\end{fminipage}

 Thus, the main question remaining is whether it is possible for the entire algorithm, including black-box queries and other elementary operations, to be efficient.

\subsubsection{Quantum Fourier transform approaches}
One natural approach is to mimic the algorithm for the Abelian HSP, which starts by computing \[ \frac{1}{\sqrt{|G|}}\sum_{x\in G} \ket{x} \ket{f(x)} \] and noticing that this equals
\[ \sum_{y \in f(G)} \frac{1}{\sqrt{|f(G)|}}\left( \frac{1}{\sqrt{|K|}} \sum_{x \in f^{-1}(y)} \ket{x} \right) \ket{y} .\]
For convenience, we suppose we measure the second register (it suffices just to trace out the 2nd register) and get some random outcome $y$, and thereby project the first register into the state
\[ \ket{a+K} = \frac{1}{\sqrt{|K|}} \sum_{x \in K} \ket{a+x}  \] where $a$ is any element of $G$ such that $f(a)=y$. In other words, as in the Abelian HSP algorithm, the first register is an equal superposition of all the elements in some coset of $K$.

The question is: how do we extract information about $K$ given a random coset state of $K$?
In the Abelian case, a quantum Fourier transform of the coset state allowed us to sample elements that were orthogonal to $K$, which we illustrated in the case of Simon's algorithm.

A more general way of looking at the quantum Fourier transform of an Abelian group $G = \mathbb{Z}_{N_1} \times \mathbb{Z}_{N_2} \times \cdots \times \mathbb{Z}_{N_l}$ is in terms of representation theory.
The Abelian group $G$ can be represented by homomorphisms $\rho$ that maps $G$ to the complex numbers.
There are in fact $|G|$ such homomorphisms, in a natural one-to-one correspondence with the elements of $G$.
For each $g = (a_1, a_2, \ldots, a_l) \in G$, define $\rho_g$ to be the homomorphism that maps any $h = (b_1, b_2 , \ldots, b_l) \in G$ according to \[ \rho_g(h) = e^{2 \pi i \sum_{i=1}^l \frac{a_i b_i}{N_i}} .\]
Using these definitions, the quantum Fourier transform maps
\[ \ket{g} \mapsto \frac{1}{\sqrt{|G|}} \sum_{h \in G} \rho_g(h) \ket{h} .\]

It is easy to verify that the quantum Fourier transform $QFT_{N_1} \otimes QFT_{N_2} \otimes \ldots \otimes QFT_{N_l}$ maps a coset state of a subgroup $K$ to a superposition of labels $h$ where $K$ is in the kernel of $\rho_h$, that is, $\rho_h(k) = 1$ for all $k \in K$. Thus, after applying the quantum Fourier transform, one will only measure labels $h = (b_1, b_2, \ldots, b_l)$ such that
\[ \rho_g(h) = e^{2 \pi i \sum_{i=1}^l \frac{a_i b_i}{N_i}} = 1 \] which generalizes the notion of orthogonality defined in the explanation of Simon's algorithm in a very natural way, since it means $\sum_{i=1}^l \frac{a_i b_i}{N_i} = 0 \mod 1$. This gives us a linear equation that must be satisfied by all $h = (b_1, b_2, \ldots, b_l) \in K$, and thus after $l + O(1)$ such random equations are collected, we can efficiently find generators for $K$ by solving a linear system of equations.

One can also define representations for finite non-Abelian groups $G$, except in order to fully capture the structure of $G$, we allow homomorphisms $\rho$ to invertible matrices over $\mathbb{C}$ (in the Abelian case, we only need $1 \times 1$ matrices). For any finite group $G$ one can define a finite set of such homomorphisms $\rho_1, \rho_2, \ldots , \rho_k$ that map elements of $G$ to unitary matrices of dimension $d_1 \times d_1$, $d_2 \times d_2, \ldots, d_k \times d_k$, respectively, with the property that $\sum_i d_i^2 = |G|$, and any other such representation is equivalent to a representation that can be factored into a collection of some number of these representations. More precisely, any representation $\rho: G \rightarrow M_{d\times d} (\mathbb{C})$ has the property that there exists some invertible $P \in M_{d\times d} (\mathbb{C})$, and a list of $\rho_{j_1}, \rho_{j_2},\ldots $ such that $\sum_i d_i = d$ and for every $g\in G$
\[ P \rho(g) P^{-1} = \oplus_i \rho_{j_i}(g) \] (that is, $P \rho(g) P^{-1}$ is block-diagonal, with the matrices $\rho_{j_i}(g)$ along the diagonals).
Furthermore, the representations $\rho_j$ are irreducible in the sense that they cannot be decomposed into the sum of two or more smaller representations in this way. They are also unique up to conjugation.

Since $\sum_i d_i^2 = |G|$, it is possible to define a unitary transformation from
$G$ to the vector space spanned by the labels of all the entries $\rho_i(j,k)$ of these irreducible representations $\rho_i$.
In particular, we can map \[ \ket{g} \mapsto \sum_i \sqrt{\frac{d_i}{|G|}} \sum_{0 \leq j,k \leq d_i} \rho_i(j,k)(g)\ket{i, j, k} \]
where $\rho_i(j,k)(g)$ is the value of the $(j,k)$ entry in the matrix $\rho_i(g)$.
Such a mapping is unitary and is called a quantum Fourier transform for the group $G$. There is freedom in the specific choice of $\rho_i$ within the set of unitary representations that are conjugate to $\rho_i$, so there is more than one quantum Fourier transform.

There has been much study of such quantum Fourier transforms for non-Abelian groups, which are sometimes possible to implement efficiently \cite{Bea97, PRB99, RB98, HRT00, MRS06}, but efficient constructions are not known in general. It appears they are of limited use in solving the non-Abelian hidden subgroup, except in special cases \cite{Ett98, RB98, HRT00, GSVV01} such as when $K$ is a normal subgroup of $G$.

In the next sections we discuss several other lines of attack on the non-Abelian hidden subgroup that have yielded some partial progress on the problem.

\subsubsection{``Pretty Good Measurements''}

A natural approach to solving the non-Abelian hidden subgroup problem is to prepare several instances of a random coset state for the hidden subgroup $K$, and then try to determine what $K$ is. More precisely, after preparing
\[ \sum_{x \in G} \ket{x} \ket{f(x)} = \sum_{y + K \in G/K} \ket{y+K} \ket{f(y)} \] and discarding the second register we are left with the mixed state
\[ \rho_K = \sum_{y+K \in G/K} \ket{y+K}\bra{y+K} .\]
Thus one could try to implement or approximate the optimal quantum measurement for identifying the mixed states $\rho_K$, over all possible hidden subgroups $K \leq G$. Furthermore, one could sample the state $\rho_K$ a polynomial number of times $t$, and try to guess $K$ given $\rho_K \otimes \rho_K \otimes \ldots \otimes \rho_K = \rho_K^{t}$.

Holevo \cite{holevo} determined the optimal measurement for the following general state distinguishability problem. Given $\rho \in \{ \rho_j \}$, output a label $m$ such that the probability that $\rho = \rho_m$ is maximum. Let $p_j$ denote the probability that $\rho = \rho_j$.
Holevo proved that the maximum probability of guessing the correct input state $\rho$ is achieved by a POVM with elements $\{G_j\}$ satisfying the following conditions:
$\sum_i p_i \rho_i G_i = \sum_i p_i G_i \rho_i$ and $\sum_i p_i \rho_i G_i \geq p_j \rho_j$. However, it is not in general easy to efficiently find and implement such a POVM.

Hausladen and Wootters \cite{HW94} defined a `pretty good' measurement for distinguishing quantum states that is not necessarily optimal, but has a simpler form.
The measurement used POVM elements $G_j = T^{-\frac{1}{2}} \rho_j T^{-\frac{1}{2}}$ and, if these don't sum to the identity, also $I - \sum_j G_j$, where $T = \sum_i \rho_i$.
For the case of the dihedral hidden subgroup problem, it was shown \cite{BCD05} that the pretty good measurement is in fact optimal; however, in this case, it is still not known how to efficiently implement the pretty good measurement. However, it was later shown how to implement the pretty good measurement for the Heisengroup group HSP \cite{BCD05}.

For example, in the case of the dihedral hidden subgroup problem for $D_{2N}$, after a quantum Fourier transform on each of $n$ coset states, one can view the resulting state of $n$ registers as \[ (\ket{0} + e^{2 \pi i \frac{k_1 d}{N}} \ket{1}) \ket{k_1} \otimes (\ket{0} + e^{2 \pi i \frac{k_2 d}{N}}\ket{1}) \ket{k_2} \otimes \ldots \otimes (\ket{0} + e^{2 \pi i \frac{k_n d}{N}}\ket{1}) \ket{k_n} \] for $k_i$ selected independently and uniformly at random from $\{0,1, \ldots, N-1\}$.

The $(\ket{0} + e^{2 \pi i \frac{k_1 d}{N}}\ket{1}) \otimes (\ket{0} + e^{2 \pi i \frac{k_2 d}{N}}\ket{1}) \otimes \ldots \otimes (\ket{0} + e^{2 \pi i \frac{k_n d}{N}}\ket{1}) $ part of the state can be rewritten as
\[ \sum_r \alpha_r \ket{S_r} \]
where $r$ spans all the possible sums of the form $\sum_i {b_i k_i}$, $b_i \in \{0,1\}$ (the `subset sums'), $\ket{S_r}$ is the uniform superposition of the strings $\ket{b_1 b_2 \ldots b_n}$ that satisfy $\sum_{b_i k_i} = r$, and $\alpha_r$ is the appropriate normalization factor (i.e. $\sqrt{n_r/ 2^n}$ where $n_r$ is the number of solutions to $\sum_{b_i k_i} = r$).

The optimal measurement \cite{BCD05} (in the restricted case of order 2 subgroups) can be thought of as mapping $\ket{S_r} \mapsto \ket{r}$, performing an inverse quantum Fourier transform and then measuring a value $\tilde{d}$ which will be the guess of the value $d$ (interestingly, a similar measurement is optimal \cite{DDEMM07b} in the case of trying to optimally estimate an arbitrary phase parameter $\phi \in [0, 2 \pi)$ given the state $(\ket{0} + e^{i k_1 \phi} \ket{1}) \times (\ket{0} + e^{ i \phi k_2} \ket{1}) \times \ldots \times (\ket{0} + e^{i \phi k_n}\ket{1})$).  Note that implementing such a basis change in reverse would solve the subset sum problem (which is NP-hard). In fact, it suffices to solve the subset sum problem on average \cite{Reg04}. A nice discussion of this connection can be found in \cite{BCD05}.

 For groups that are semidirect products of an Abelian group and a cyclic group, the pretty good measurement corresponds to solving what is referred to \cite{BCD05} as a `matrix sum problem', which naturally generalizes the subset sum problem. They also show that the pretty good measurements are optimal in these cases, and similarly relate their implementation to solving the matrix sum problem to certain average-case algebraic problems. They show that the pretty good measurement can be implemented for several groups including semidirect product groups of the form $\Z_p^r \rtimes \Z_p$ for constant $r$  (when $r=2$, this is the Heisenberg group), and of the form $\Z_N \rtimes \Z_p$ with $p$ prime (which are {\it metacyclic}) and where the ratio $N/p$ is polynomial in $\log N$.

\subsubsection{``Sieving''}

Kuperberg \cite{Kup05} introduced a method for attacking the hidden subgroup problem for the dihedral group that leads to a sub-exponential algorithm.

The dihedral group $D_{N}$ is a non-Abelian group of order $2N$, which corresponds to the set of symmetries of a regular $N$-gon. It can be defined as the set of elements $\{ (x,d) \mid x \in \{0,1,2, \ldots, N-1\}, d \in \{0,1\} \}$, where $\{ (x,0) \mid x \in \{0,1,2, \ldots, N-1\} \}$ is the Abelian subgroup of $D_N$ corresponding to the rotations (satisfying $(x,0)+(y,0) = (x+y \mod N,0)$), and $\{ (0,0), (y,1) \}$ are Abelian subgroups of order $2$ corresponding to reflections. In general, $(x,0)+(y,1) = (y-x,1)$, $(y,1)+(x,0) = (x-y,1) = -((x,0)+(y,1))$.  If the hidden subgroup is a subgroup of the Abelian subgroup of order $N$, then finding the hidden subgroup easily reduces to the Abelian hidden subgroup problem.  However, there is no known efficient algorithm for finding hidden subgroups of the form $\{ (0,0), (y,1) \}$. So we can focus attention to the following restricted version of the dihedral hidden subgroup.

\begin{fminipage}{\mylength}
\textbf{\underline{Dihedral Hidden Subgroup Problem (hard case)}} \vskip 6pt
\textbf{Input:}
An integer $N$, and a black-box implementing $U_f: \ket{x,d} \ket{0} \mapsto \ket{x,d} \ket{f(x,d)}$, where $f(x_1, d_1) = f(x_2, d_1)$ if and only if $(x_1, d_1)-(x_2, d_2) \in \{(0,0), (y,1) \}$ for some $y$.

\textbf{Problem:} Find $y$.
\end{fminipage}

As mentioned earlier, the dihedral hidden subgroup problem can be efficiently reduced to the following phase estimation problem \cite{EH00}.

\begin{fminipage}{\mylength}
\textbf{\underline{Ettinger-H{\o}yer phase estimation problem for the dihedral HSP}} \vskip 6pt
\textbf{Input:}
An integer $N$, and a black-box $O_{d}$ that outputs a classical value $k \in \{0,1, \ldots, N-1\}$ uniformly at random, along with the qubit $\frac{1}{\sqrt{2}}( \ket{0} + e^{i \phi k} \ket{1})$ where $\phi = \frac{2 \pi d}{N}$, for some integer $d \in \{0, 1, 2 \ldots, N-1\}$.

\textbf{Problem:} Find $d$.
\end{fminipage}

Note that if we could sample the values $k=2^j$ for $j = 1,2, \ldots , \lceil \log N \rceil$, then the phase estimation problem can be solved directly using the quantum Fourier transform \cite{CEMM98}.

Regev designed a clever method for generating states of the form $\frac{1}{\sqrt{2}} (\ket{0} + e^{i  2^j \phi }  \ket{1})$ using $O(1)$ calls to the black-box $O_d$, given an oracle that solves the subset sum problem on average.
Kuperberg \cite{Kup05} developed a ``sieving'' method of generating states of the form $\frac{1}{\sqrt{2}} (\ket{0} + e^{i  2^j \phi } \ket{1})$ and the method was refined and improved to use less memory by Regev \cite{Reg04}.

\begin{fminipage}{\mylength}
\textbf{\underline{Quantum algorithms for the dihedral HSP}} \vskip 6pt
There exists a quantum algorithm that solves the dihedral HSP with running time in $e^{O(\sqrt{\log N})}$ and uses space in $e^{O(\sqrt{\log N})}$. There is also an algorithm with running time in $e^{O(\sqrt{\log N \log \log N})}$ and uses space in $O(\log N)$.
\end{fminipage}

\begin{fminipage}{\mylength}
\textbf{\underline{Classical algorithms for the dihedral HSP}} \vskip 6pt
The classical complexity of the dihedral hidden subgroup problem is $\Theta( \sqrt{N})$ evaluations of the function $f$.
\end{fminipage}

Similar sieving methods were applied  \cite{AMR07}  to yield a subexponential time algorithm for the HSP over the product groups $G^n$ for a fixed non-Abelian group $G$.
It has also been show that these kinds of quantum sieve algorithms will not give efficient quantum algorithms for graph isomorphism \cite{MRS07}.

\subsubsection{Other methods and results}
There are also some algorithms for solving other cases of the non-Abelian hidden subgroup problem that don't use any of the above techniques, e.g. \cite{IMS01, FIMSS03, ISS07}. These results use sophisticated classical and quantum group theoretic techniques to reduce an instance of a non-Abelian HSP to instances of HSP in Abelian groups.

One of the most recent results \cite{ISS08} shows that such a reduction is possible for {\it nil-2} groups, which are nilpotent groups of {\it class} 2. The group $G$ is nilpotent of class $n$ if the following holds. Let $A_1 = G$, and let $A_{i+1} = [A_i, G]$, for $i > 0$. A group $G$ is nilpotent if $A_{n+1} = \{ 1 \}$, for some integer $n$, and the class of a nilpotent group is the smallest positive integer $n$ for which $A_{n+1} = \{1\}$.

One of their techniques is to generalize Abelian HSP to a slightly more general problem, where the hidden subgroup $K$ is hidden by a quantum procedure with the following properties.
For every $g_1, g_2, \ldots, g_N \in G$ the algorithm maps
\begin{eqnarray*}
\ket{g_1} \ket{g_2} \ldots \ket{g_N} \ket{0} \ket{0} \ldots \ket{0} \mapsto \ket{g_1} \ket{g_2} \ldots \ket{g_N} \ket{\psi_{g_1}^1} \ket{\psi_{g_2}^2} \ldots \ket{\psi_{g_N}^N} \end{eqnarray*}
where the set of states $\{ \ket{\psi_{g}^i} \mid g \in G \}$ is a {\it hiding set} for $K$, for each $i=1,2, \ldots, N$.
A set of normalized states $\{ \ket{\psi_{g}} \mid g \in G \}$ is a {\it hiding set} for the subgroup $K$ of $G$ if

$\bullet$
If $g$ and $h$ are in the same left coset of $K$ then $ \ket{\psi_g} = \ket{\psi_h}$.

$\bullet$
If $g$ and $h$ are in different left cosets of $K$ then $\bra{\psi_g} \psi_j \rangle = 0$.

Generators for the subgroup $K$ of $G$ can be found in time polynomial in $\log |G|$ using a quantum hiding procedure with $N \in O(\log |G|)$.
They find a series of non-trivial reductions of the standard HSP in nil-2 groups to instances of the Abelian HSP with a quantum hiding function.

\subsection{Lattice and Number Field Problems} \label{lattice}

The Abelian Hidden Subgroup Problem also works for finitely generated groups $G$. We can thus define the hidden subgroup problem on $G = \mathbb{Z} \times \mathbb{Z} \times \ldots \times \mathbb{Z} = \mathbb{Z}^n$.
The hidden subgroup $K$ will be generated by some $n$-tuples in $\mathbb{Z}^n$. We can equivalently think of $G$ as a lattice and $K$ as a sublattice. The function $f : G \rightarrow X$, for some finite set $X$, that satisfies $f(\x) = f(\y)$ if and only if $\x - \y \in K$, can be thought of as hiding the sublattice $K$.

By generalizing the problem to hiding sublattices of $\mathbb{R}^n$, one can solve some interesting and important number theoretic problems. The solutions in these cases were not a simple extension of the Abelian hidden subgroup algorithm.

Hallgren \cite{Hal02, Hal07} found a quantum algorithm for finding the integer solutions $x,y$ to Pell's equation $x^2 - d y^2 =1$, for any fixed integer $d$.
He also found an efficient quantum algorithm for the Principal Ideal Problem, and later generalized it to computing the unit group of a number field of constant degree \cite{Hal05}. Solving Pell's equation is known to be at least as hard as factoring integers. We don't have room to introduce all the necessary definitions, but we'll sketch some of the main ideas.

A number field $F$ can be defined as a subfield $Q(\theta)$ of the complex numbers that is generated by the rationals $Q$ together with a root $\theta$ of an irreducible polynomial with rational coefficients; the degree of $F$ is the degree of the irreducible polynomial. The ``integers'' of $F$ are the elements of $F$ that are roots of {\it monic} polynomials (i.e. polynomials with leading coefficient equal to $1$, such as $x^2 + 5x+ 1$). The integers of $F$ form a ring, denoted $\mathcal{O}$. One can define a parameter $\Delta$ called the {\it discriminant} of $F$ (we won't define it here), and an algorithm is considered efficient if its running time is polynomial in $\log \Delta$ and $n$. The unit group of the ring $F$, denoted $\mathcal{O}^*$, is the set of elements $\alpha$ in $\mathcal{O}$ that have an inverse $\alpha_{-1} \in \mathcal{O}$. ``Computing'' the unit group corresponds to finding a description of a system of ``fundamental'' units, $\epsilon_1, \epsilon_2, \ldots, \epsilon_r$ that generate $\mathcal{O}^*$ in the sense that every unit $\epsilon \in \mathcal{O}^*$ can be written as $\epsilon = \mu \epsilon_1^{k_1} \epsilon_2^{k_2} \ldots \epsilon_r^{k_r}$ for some $k_1, k_2, \ldots, k_r \in \Z$ and some root of unity $\mu$. However, in general, an exact description of a fundamental unit requires an exponential number of bits. There are are some finite precision representations of the elements of the unit group, such as the ``Log'' embedding into $\R^r$. This representation describes a unit $\alpha$ by an element of $\R^r$ where some finite precision representation of each coordinate suffices. This Log representation of the unit group $\mathcal{O}^*$, corresponds to a sublattice $L = \mathcal{O}^*$ of $\R^r$. Hence, we have a relationship between several important computational number field problems, and lattice problems.

By the above correspondence between $\mathcal{O}^*$ and the lattice $L \subset \R^r$, we can formulate \cite{Hal05, SV05} the problem of computing the unit group as the problem of finding elements that approximate generators of the sublattice $L$ of $\R^r$.
One important non-trivial step is defining a function $f : \R^r \rightarrow X$ (for some infinite set $X$) such that $f(x) = f(y)$ if and only if $x-y \in L$ as well as appropriate discrete approximations to this function. The definition of these functions involves substantial knowledge of algebraic number theory, so we will not describe them here.

By designing quantum algorithms for approximating generators of the lattice $L$, \cite{Hal05, SV05} one can find polynomial time algorithms for computing the unit group $\mathcal{O}^*$ of an algebraic number field $F = \mathbb{Q}(\theta)$.

A corollary of this result, is a somewhat simpler solution to the Principal Ideal problem (in a constant degree number field) that had been found earlier by Hallgren \cite{Hal02}. An ideal $\mathcal{I}$ of the ring $\mathcal{O}$ is a subset of elements of $\mathcal{O}$ that is closed under addition, and is also closed under multiplication by elements of $\mathcal{O}$. An ideal $\mathcal{I}$ is {\it principal} if it can be generated by one element $\alpha \in \mathcal{I}$; in other words $\mathcal{I} = \alpha \mathcal{O} = \{ \alpha \beta \mid \beta \in \mathcal{O} \}$. The principal ideal problem is, given generators for $\mathcal{I}$, to decide if $\mathcal{I}$ is principal, and if it is, to find a generator $\alpha$.

As mentioned in section \ref{AHSP}, the tools developed can also be applied to find unique (quantum) representatives of elements of the class group for constant degree number fields \cite{Hal07} (assuming the Generalized Riemann Hypothesis), and thus allow for the computation of the class group in these cases.

\subsection{Hidden Non-Linear Structures}

Another way to think of the Abelian Hidden Subgroup Problem is as an algorithm for finding a hidden linear structure within a vector space. For simplicity, let's consider the Abelian HSP over the additive group $G=\mathbb{F}_q \times \mathbb{F}_q \times \ldots \times \mathbb{F}_q = \mathbb{F}_q^n$, where $q = p^m$ is a prime power. The elements of $G$ can also be regarded as a vector space over $\mathbb{F}_q$. A hidden subgroup $H \leq G$ corresponds to a subspace of this vector space and its cosets correspond to parallel affine subspaces or {\it flats}. The function $f$ is constant on these linear structures within the vector space $\mathbb{F}_q^n$.

A natural way to generalize this \cite{CSV07} is to consider functions that are constant on sets that are defined by non-linear equations.
One problem they study is the {\it hidden radius problem}. The circle of radius $r \in \mathbb{F}_q$ centred at $t = (t_1, t_2, \ldots, t_n) \in \mathbb{F}_q^n$ is the set of points $x = (x_1, x_2, \ldots, x_n) \in \mathbb{F}_q^n$ that satisfy $\sum_i (x_i - t_i)^2 = r$.  The point $x$ on a circle centered at $t$ will be represented as either $(x, \sigma(s))$, where $s = x-t$ and $\sigma(s)$ is a random permutation of $s$, or as $(x, \tau(t))$ where $\tau(t)$ is a random permutation of $t$.
%The ``encryptions'' are random one-to-one functions into some finite range.
We define $f_1$ and $f_{-1}$ be the functions satisfying $f_1(x, \sigma(s)) = \tau(t)$ and $f_{-1}(x, \tau(t)) = \sigma(s)$.
%MM check this definition

\begin{fminipage}{\mylength}
\textbf{\underline{Hidden Radius Problem}} \vskip 6pt
\textbf{Input:}
A black box $U_{f_1}$ that implements $\ket{x} \ket{\sigma(s)} \ket{0} \mapsto \ket{x} \ket{\sigma} \ket{f_1(x,\sigma(s))}$, and a black box $U_{f_{-1}}$ that implements
$\ket{x} \ket{\tau(t)} \ket{0} \mapsto \ket{x} \ket{\tau(t)} \ket{f_{-1}(x,\tau(t))}$.

\textbf{Problem:} Find $r$.
\end{fminipage}

\begin{fminipage}{\mylength}
\textbf{\underline{Quantum Algorithms for the Hidden Radius Problem}} \vskip 6pt
For odd $d$, there is a bounded error quantum algorithm that makes $\poly \log(q)$ queries to $U_{f_1}$ and $U_{f_{-1}}$ and finds $r$. However, there is no known polynomial bound on the non-query operations.

There is a bounded-error quantum algorithm that also makes $\polylog (q)$ operations in total, and determines $\chi(r)$, where $\chi(r) = 1$ if $r$ is a quadratic residue (that is, $r = u^2$ for some $u \in \mathbb{F}_q$) and $0$ otherwise.
\end{fminipage}

\begin{fminipage}{\mylength}
\textbf{\underline{Classical Algorithms for the Hidden Radius Problem}} \vskip 6pt
It was shown in \cite{CSV07} that the expected number of queries needed to be able to guess any bit of information about $r$ correctly with probability greater than $\frac{1}{2} + \frac{1}{\poly (d \log q)}$ is exponential in $d \log q$.
\end{fminipage}

A number of other black-box problems of this kind were defined in \cite{CSV07} with quantum algorithms that are exponentially more efficient than any classical algorithm, in some cases just in terms of query complexity, and other times in terms of all operations.
These problems fit into the frameworks of {\it shifted subset} problems and {\it hidden polynomial} problems. They use a variety of non-trivial techniques for these various problems, including the quantum Fourier transform, quantum walks on graphs, and make some non-trivial connections to various {\it Kloosterman sums}. Further work along these lines has been done in \cite{DDW08}.

\subsection{Hidden shifts and translations}

There have been a variety of generalizations of the hidden subgroup problem to the problem of finding some sort of hidden shift or translation.

Grigoriev \cite{Gri97} addressed the problem of the shift-equivalence of two polynomials (the self-shift-equivalence problem is a special case of the Abelian Hidden Subgroup Problem). Given two
polynomials $P_1, P_2$ in $l$ variables $X_1,X_2,\ldots,X_l$ over
$\mathbb{F}_q$ (the finite field with $q$ elements), does there exist an element $(a_1,a_2,\ldots,a_l)\in\mathbb{F}_q^l$ such that
$P_1(X_1-a_1,X_2-a_2,\ldots,X_l-a_l) = P_2(X_1, X_2, \ldots, X_l)$.
More generally, if there is a group $G$ ($ \mathbb{F}_q^l$ in this case) acting on a set $X$ (the set of polynomials in $l$ variables over $\mathbb{F}_q$ in this case), one can ask if two elements $x,y \in X$ are in the same orbit of the action of $G$ on $X$ (that is, if there is a $g \in G$ such that $g(x) = y$). In general, this seems like a hard problem, even for a quantum computer.

The dihedral hidden subgroup problem \cite{EH00} is a special case of the {\it hidden translation problem} \cite{FIMSS03}, where there is a finite group $G$, with unique representation of the elements of $G$, and two injective functions $f_0$ and $f_1$ from $G$ to some finite set $X$.
\begin{fminipage} {\mylength}
\textbf{\underline{Hidden Translation Problem:}} \vskip 6pt
\textbf{Input:}

Two black boxes $U_{f_0}$ and $U_{f_1}$ that, for any $x \in G$, implement the maps $U_{f_0}: \ket{x}\ket{0} \mapsto \ket{x} \ket{f_0(x)}$ and $U_{f_1}: \ket{x}\ket{0} \mapsto \ket{x} \ket{f_1(x)}$.

A promise that $f_1(x) = f_0(ux)$, for some $u \in G$.

\textbf{Problem:} Find $u$.
\end{fminipage}

The same problem expressed with additive group notation has been called the {\it hidden shift problem}, and instances of this problem were solved efficiently by van Dam, Hallgren and Ip \cite{DHI03}. For example, they find an efficient solution in the case that $f_1(x) = \chi(x+s)$  where $f_0 = \chi$ is a multiplicative character function over a finite field, which implies a method for breaking a class of ``algebraically homomorphic'' cryptosystems. They also describe a more general {\it hidden coset problem}.

In \cite{FIMSS03} it is shown how to solve the hidden translation problem in $G = \Z_p^n$ in polynomial time, and then show how to use this to solve the problem for any group that they call ``smoothly solvable''. Let us briefly define what this means.

For any two elements $g,h$ of a group, we define their commutator, denoted $[g,h]$ to be $[g,h] = g^{-1} h^{-1} g h$, and for any two subgroups $H, K \leq G$ we define $[H,K]$ to be the (normal) subgroup of $G$ generated by all the commutators $[h,k]$ where $h \in H, k \in K$. The {\it derived subgroup} of a group $G$ is $G^{(1)} = [G,G]$.
In general, we define $G^{(0)}  = G, G^{(n+1)} = [G^{(n)}, G^{(n)}]$, for $n \geq 1$.
A group $G$ is solvable if $G^{(n)} = \{ 1 \}$, the trivial group, for some positive integer $n$, and the series of subgroups is called the {\it derived series} of $G$.
A group $G$ is called smoothly solvable if $m$ is bounded above by a constant, and if the factor groups $G^{(j+1)}/G^{(j)}$ are isomorphic to a direct product of a group of bounded exponent (the exponent of a group is the smallest positive integer $r$ such that $g^r = 1$ for all $g$ in the group) and a group of size polynomial in $\log |G|$.

The algorithm for smoothly solvable groups works by solving a more general {\it orbit coset} problem,
 for which they prove a ``self-reducibility'' property. In particular, orbit coset problem for a finite group $G$ is reducible to the orbit coset problem in $G/N$ and $N$, for any solvable normal subgroup $N$ of $G$.

\begin{fminipage} {\mylength}
\textbf{\underline{Quantum Algorithms for the Hidden Translation Problem:}} \vskip 6pt

For groups $G = \Z_p^n$, the hidden translation problem can be solved with bounded probability using $O(p(n+p)^{p-1})$ queries and $(n+p)^{O(p)}$ other elementary operations.

In general, for {\it smoothly solvable} groups $G$, the hidden translation problem can also be solved with a polynomial number of queries and other elementary operations. Another consequence of the tools they developed is a polynomial time solution to the hidden subgroup for such smoothly solvable groups.

\end{fminipage}

\begin{fminipage} {\mylength}
\textbf{\underline{Classical Algorithms for the Hidden Translation Problem:}} \vskip 6pt

In general, including the case $G = \Z_p^n$, the hidden translation problem requires $\Omega(\sqrt{|G|})$ queries on a classical computer.
\end{fminipage}

Another natural generalization of the hidden translation or hidden shift problem and the Abelian Hidden Subgroup Problem is the {\it generalized hidden shift problem} introduced in \cite{CD07}. There is a function $f: \{0,1,2,\ldots, M-1\} \times \Z_N \rightarrow X$ for some finite set $X$, with the property that for a fixed $b \in \{0,1,\ldots, M-1\}$, the mapping $x \mapsto f(b,x)$ is one-to-one, and there is some hidden value $s \in \Z_N$ such that $f(b,x) = f(b+1, x+s)$ for all $b \in \{0,1, \ldots, M-2\}$.  Note that for $M=2$, this is equivalent to the dihedral hidden subgroup problem for the group $D_{N}$, and for $M=N$, this problem is equivalent to the Abelian Hidden Subgroup Problem for the hidden subgroup $\langle (1,s)\rangle$ of the group $\Z_N \times \Z_N$.

\begin{fminipage} {\mylength}
\textbf{\underline{Generalized Hidden Shift Problem:}} \vskip 6pt
\textbf{Input:}
Positive integers $M$ and $N$.

A black boxes $U_{f}$ that maps $\ket{b,x}\ket{0} \mapsto \ket{b,x} \ket{f(b,x)}$ for all $b \in \{0,1, \ldots, M-1\}$ and $x \in \Z_N$, where $f$ satisfies the properties defined above.

\textbf{Problem:} Find $s$.
\end{fminipage}

\begin{fminipage} {\mylength}
\textbf{\underline{Quantum Algorithms for the Generalized Hidden Shift Problem:}} \vskip 6pt

There is a quantum algorithm that, for any fixed $\epsilon > 0$, and $M \geq N^{\epsilon}$, solves the generalized hidden shift problem in time polynomial in $\log N$.

\end{fminipage}

The algorithm uses a ``pretty good measurement'' that involves solving instances of the following matrix sum problem. Given $\x \in \Z_N^k$ and $w \in Z_N$ chosen uniformly at random, find $\mathbf{b} \in \{0,1, \ldots, M-1 \}^k$ such that $\sum b_i x_i = w \mod N$. Note how this generalizes the subset sum problem, which was shown to be related to the dihedral hidden subgroup problem \cite{Reg04}. While there is no efficient solution known for small $M$ (even an average case solution suffices for the dihedral HSP), for $M \geq N^{\epsilon}$, Lenstra's integer programming algorithm allows for an efficient solution to the matrix sum problem.

\begin{fminipage} {\mylength}
\textbf{\underline{Classical Algorithms for the Generalized Hidden Shift Problem:}} \vskip 6pt

Any classical algorithm requires $\Omega(\sqrt{N})$ evaluations of the function $f$.

\end{fminipage}

\subsection{Other related algorithms}

There are a variety of other problems that aren't (as far as we know) generalizations of the hidden subgroup problem, and arguably deserve a separate section. We'll mention them here since various parts of the algorithms for these problems use techniques related to those discussed in one of the other subsections of this section.

Van Dam and Seroussi \cite{DS02} give an efficient quantum algorithm for estimating Gauss sums.
Consider a finite field $\mathbb{F}_{p^r}$ (where $p$ is prime, and $r$ is a positive integer).  The multiplicative characters are homomorphisms of the multiplicative group, $\mathbb{F}_{p^r}^*$,  to the complex numbers $\mathbb{C}$, and also map $0 \mapsto 0$. Each multiplicative character can be specified by an integer $\alpha \in \{0,1, \ldots, p^r-2 \}$ by defining $\chi_{\alpha}(g^j) = \xi^{\alpha j}$, where $g$ is a generator for $\mathbb{F}_{p^r}^*$ and $\xi = e^{2 \pi i/(p^r-1)}$.

The additive characters are homomorphisms of the additive group of the field to $\mathbb{C}$, and can be specified by a value $\beta \in \mathbb{F}_{p^r}$ according to $e_{\beta}(x) = \zeta^{Tr(\beta x)}$, where $Tr(y) = \sum_{j=0}^{r-1} y^{p^j}$ and $\zeta = e^{2 \pi i/p}$.

The Gauss sum $G(\mathbb{F}_{p^r}, \chi_{\alpha}, e_{\beta})$ is defined as
\[ G(\mathbb{F}_{p^r}, \chi, e_{\beta}) = \sum_{x \in \mathbb{F}_{p^r}} \chi(x) e_{\beta}(x) .\]

It is known that the norm of the Gauss sum is $|G(\mathbb{F}_{p^r}, \chi, e_{\beta})| = \sqrt{p^r}$, and thus the hard part is determining, or approximating, the parameter $\gamma$ in the equation $G(\mathbb{F}_{p^r}, \chi, e_{\beta}) = e^{i \gamma} \sqrt{p^r}$.

\begin{fminipage} {\mylength}
\textbf{\underline{Gauss Sum Problem for Finite Fields:}} \vskip 6pt

{\bf Input:}
A prime number $p$, positive integer $r$ and a standard specification of $\mathbb{F}_{p^r}$ (including a generator $g$).

A positive integer $\alpha \in \{0,1, \ldots, p^r - 2 \}$.

An element $\beta \in \mathbb{F}_{p^r}$.

A parameter $\epsilon$, $0 < \epsilon < 1$.

{\bf Problem:}
Output an approximation, with error at most $\epsilon$, to $\gamma$ in the equation $G(\mathbb{F}_{p^r}, \chi_{\alpha}, e_{\beta}) = e^{i \gamma} \sqrt{p^r} $.
\end{fminipage}

One noteworthy feature of this problem is that it is not a black-box problem.

\begin{fminipage} {\mylength}
\textbf{\underline{Quantum Algorithms for the Finite Field Gauss Sum Problem:}} \vskip 6pt
There is a quantum algorithm running in time $O(\frac{1}{\epsilon} \polylog (p^r) )$ that outputs a value $\tilde{\gamma}$ such that $|\gamma - \tilde{\gamma}| < \epsilon$ with probability at least $\frac{2}{3}$.
\end{fminipage}

\begin{fminipage} {\mylength}
\textbf{\underline{Classical Complexity of the Finite Field Gauss Sum Problem:}} \vskip 6pt
It was shown that solving this problem is at least as hard as the discrete logarithm problem in the multiplicative group of $\mathbb{F}_{p^r}$ (see Section \ref{DLP-section}).
\end{fminipage}

Various generalizations of this problem were also studied in \cite{DS03}.
Other examples include \cite{DS08} which studies the problem of finding solutions to equations of the form $a f^x + b g^y = c$, where $a,b,c,f,g$ are elements of a finite field, and $x,y$ are integers.

\section{Quantum walk algorithms} \label{walks}

Quantum walks, sometimes called quantum random walks, are quantum analogues of (classical) random walks, which have proved to be a very powerful algorithmic tool in classical computer science.
The quantum walk paradigm is still being developed.
For example, the relationship between the continuous time and discrete time models of quantum walks is still not fully understood. In any case, the best known algorithms for several problems are some type of quantum walk.

Here we restrict attention to walks on discrete state spaces. Because of the Strong Church-Turing thesis, we expect that any practical application of a walk on a continuous state space will have an efficient (up to polynomial factors) simulation on a discrete system.

In general, any walk algorithm (classical or quantum), consists of a discrete state space, which is usually finite in size, but sometimes infinite state spaces are also considered when it is convenient to do so.
The state space is usually modeled as being the vertices of a graph $G$, and the edges of the graph denote the allowed transitions. In classical discrete time walks, the system starts in some initial state, $v_i$. Then at every time step the system moves to a random neighbour $w$ of the current vertex $v$, according to some probability distribution $p(v,w)$.  Let $M$ denote the matrix where the $(v,w)$ entry is $p(v,w)$. Let $\bf{v_0}$ be the column vector with the value $p_i$ in the $i$th position, where $p_i$ is the probability that the initial vertex is $v_i$. Then the vector ${\bf{v_t}} = M^t {\bf{v_0}}$ describes the probability distribution of the system after $t$ time steps after starting in a state described by the probability distribution ${\bf v_0}$.

The walks are usually analyzed as abstract walks on a graph. In practice, the vertices are representing more sophisticated objects.
For example, suppose one wishes to solve a 3-SAT formula $\Phi$ on $n$ Boolean variables. One could define a random walk on the $2^n$ possible assignments of the Boolean variables. So the vertices of the graph would represent the $2^n$ Boolean strings of length $n$.
One could start the walk on a random vertex (which corresponds to a random assignment of the $n$-Boolean variables).  At every step of the walk, if the current vertex $v$ corresponds to a satisfying assignment, then $p(v,v)=1$ and the walk should not leave the vertex. Otherwise, a random clause should be picked, and one of the variables in that clause should be picked uniformly at random and flipped. This implicitly defines a probability distribution $p(v,w)$.

In a quantum walk, instead of just having classical probability distributions of the vertices $v_i \in V(G)$, one can have superpositions $\sum_{v_i \in V(G)} \alpha_i \ket{v_i}$ , and more generally any quantum mixed state of the vertices. If we restrict to unitary transitions, then there is a unitary matrix $U$ that contains the transition amplitudes $\alpha(v,w)$ of going from vertex $v$ to vertex $w$, and if the systems starts in initial state $\ket{\psi_0}$, then after $k$ time steps the state of the system is $U^k \ket{\psi_0}$. These unitary walks are not really ``random'' since the evolution is deterministic. More generally, the transition function could be a completely positive map $\mathcal{E}$, and if the system starts in the initials state $\rho = \ket{\psi_0} \bra{\psi_0}$, then after $t$ time steps the state of the system will be $\mathcal{E}^t(\rho)$.

One cannot in general define a unitary walk on any graph \cite{Sev03},
 however if one explicitly adds a ``coin'' system of dimension as a large as the maximum degree $d$ of the vertices (i.e. the new state space consists of the states $\ket{v_i} \ket{c}$, $v_i \in V(G)$ and $c \in \{0,1, \ldots, d-1\}$) then one can define a unitary walk on the new graph one would derive from the combined graph-coin system. In particular, the state of the coin system indicates to which neighbour of a given vertex the system should evolve. More generally, one can define a unitary walk on states of the form $(v_i,v_j)$, where $\{v_i, v_j \}$ is an edge of $G$.

A continuous version of quantum walks was introduced by Farhi and Gutmann \cite{FG98}.
  The idea is to let the adjacency matrix of the graph be the Hamiltonian driving the evolution of the system. Since the adjacency matrix is Hermitian, the resulting evolution will be unitary. The reason such a unitary is possible even for a graph where there is no unitary discrete time evolution is that in this continuous time Hamiltonian model, for any non-zero time evolution, there is some amplitude with which the walk has taken more than one step.

In classical random walks, one is often concerned with the ``mixing time'', which is the time it takes for the system to reach its equilibrium distribution. In a purely unitary (and thus reversible) walk, the system never reaches equilibrium, but there are alternative ways of arriving at an effective mixing time (e.g. averaging over time). In general, quantum walks offer at most a quadratically faster mixing. Another property of random walks is the ``hitting time'', which is the time it takes to reach some vertex of interest. There are examples where quantum walks offer exponentially faster hitting times.

The study of what are essentially quantum walks has been around for decades, and the algorithmic applications have been developed for roughly 10 years. Much of the early algorithmic work developed the paradigm and discovered the properties of quantum walks on abstract graphs, such as the line or circle, and also on general graphs (e.g. \cite{AAKV01, ABNVW01}).
 There have also been applications to more concrete computational problems, and we will outline some of them here.

\begin{fminipage}{\mylength}
\textbf{\underline{Element Distinctness Problem}}
\vskip 6pt
{\bf Input:}
A black-box $U_f$ that maps $\ket{i}\ket{b} \mapsto \ket{i} \ket{b \oplus f(i)}$ for some function $f: \{0,1,\ldots, N-1\} \rightarrow \{0,1,\ldots, M\}$.

{\bf Problem:} Decide whether there exist inputs $i$ and $j$, $i \neq j$, such that $f(i) = f(j)$.
\end{fminipage}

Prior to the quantum walk algorithm of Ambainis, the best known quantum algorithm used $O(N^{\frac{3}{4}})$ queries.
\begin{fminipage}{\mylength}
\textbf{\underline{Quantum Algorithms for Element Distinctness Problem}}
\vskip 6pt
The quantum walk algorithm in \cite{Amb04a} uses $O(N^{\frac{2}{3}})$ evaluations of $U_f$, $O(N^{\frac{2}{3}} \polylog N)$ non-query operations and $O(N^{\frac{2}{3}}\polylog N)$ space.
\end{fminipage}
%MM todo maybe RudolphGrover discusses classical postprocessing

\begin{fminipage}{\mylength}
\textbf{\underline{Classical Algorithms for Element Distinctness}}
\vskip 6pt
A classical computer requires $N - O(1)$ applications of $U_f$ in order to guess correctly with bounded error for worst-case instances of $f$.
\end{fminipage}

As is often the case with classical random walk algorithms, the graph is only defined implicitly, and is usually exponentially large in the size of the problem instance. For the element distinctness algorithm, the graph is defined as follows.
The vertices are subsets of $\{1,2, \ldots, N\}$ of size $\lceil N^{\frac{2}{3}} \rceil$.
Two vertices are joined if the subsets differ in exactly two elements.
A detailed description and analysis of this walk is beyond the scope of this survey.

Szegedy \cite{Sze04} extended the approach of Ambainis to develop a powerful general framework for quantizing classical random walks in order to solve search problems. Suppose we wish to search a solution space of size $N$ and there are $\epsilon N$ solutions to $f(x)=1$.
Furthermore, suppose there is a classical random walk with transition matrix $M$, with the property that $p(v,w) = p(w,v)$ (known as a `symmetric' walk). It can be shown that the matrix $M$ has maximum eigenvalue $1$, and suppose the next highest eigenvalue is $1-\delta$, for $\delta > 0$.
The classical theory of random walks implies the existence of a bounded-error classical random walk search algorithm with query complexity in $O(\frac{1}{\delta \epsilon})$. Szegedy developed a ``$\sqrt{\delta \epsilon}$-rule'' that gives a quantum version of the classical walk with query complexity in $O(\frac{1}{\sqrt{\delta \epsilon}})$.  This technique was generalized further in \cite{MNRS07} and summarized nicely in \cite{San08}.

Quantum walk searching has been applied to other problems such as triangle-finding \cite{MSS07}, commutativity testing \cite{MN07}, matrix product verification \cite{BS06}, associativity testing when the range is restricted \cite{DT07}, and element k-distinctness \cite{Amb04a}. A survey of results in quantum walks can be found in \cite{Amb03, Amb04b, Kem03, San08}.

\subsection{Continuous time quantum walk algorithms} \label{continuous-walks}

In this section we describe two very well-known continuous time quantum walk algorithms. The first algorithm \cite{CCDFGS03} illustrates how a quantum walk algorithm can give an exponential speed-up in the black-box model.

A problem instance of size $n$ corresponds to an oracle $O_{G_n}$ that encodes a graph $G_n$ on $O(2^n)$ vertices in the following way. The graph $G_n$ is the graph formed by taking $2$ binary trees of depth $n$, and then ``gluing'' the two trees together by adding edges that create a cycle that alternates between the leaves of the first tree (selected at random) and the leaves of the second tree (selected at random). The two root vertices are called the ``ENTRANCE'' vertex, labelled with some known string, say, the all zeroes string $000 \ldots 0$ of length $2n$, and ``EXIT'' vertex, which is labelled with a random string of length $2n$. The remaining vertices of the graph are labelled with distinct random bit strings of length $2n$. The oracle $O_{G_n}$ encodes the graph $G_n$ in the following way. For $\ket{x}$ where $x \in \{0,1\}^{2n}$ encodes a vertex label of $G_n$, $O_{G_n}$ maps $\ket{x}\ket{00 \ldots 0}$ to $\ket{x} \ket{n_1(x), n_2(x), n_3(x)}$ where $n_1(x), n_2(x), n_3(x)$ are the labels of the neighbours of $x$ in any order (for the exit and entrance vertex, there will only be two distinct neighbours).

\begin{figure}[h!]
\begin{center}
\epsfig{file=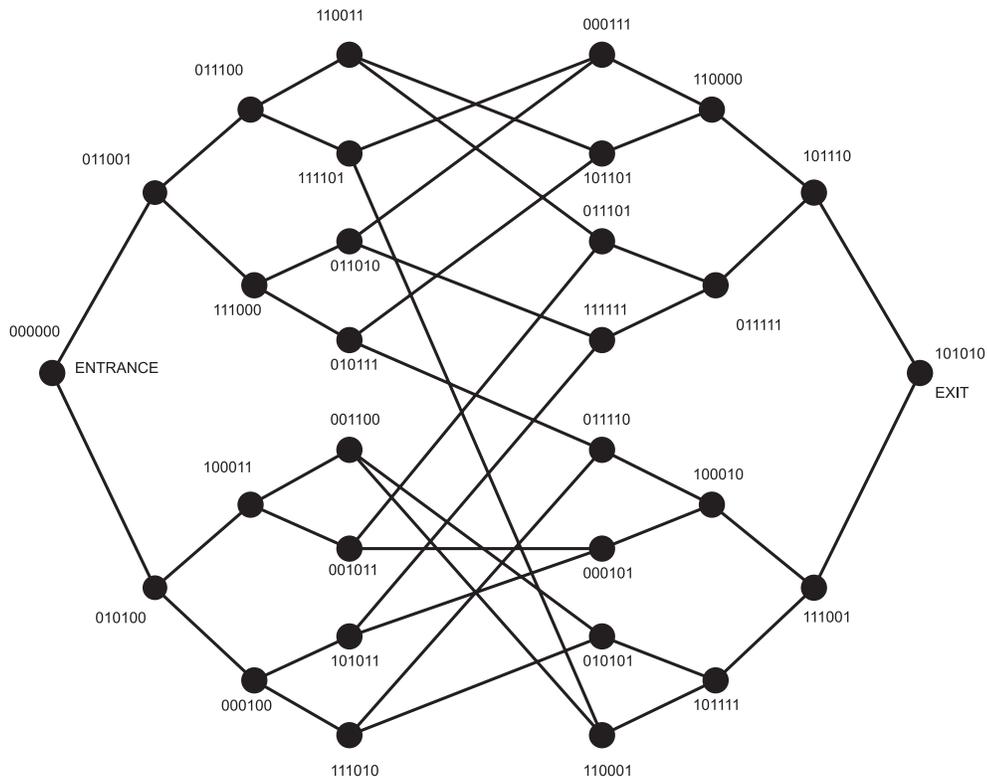,width=13cm}
\end{center}
\caption{This is an example of a ``glued-trees'' graph with random labellings for all vertices (except the ``ENTRANCE'' vertex). The goal is to find the label of the ``EXIT'' vertex (in this case, it is $101010$), given a black-box that outputs the vertex labels of the neighbours of a given vertex label.}
\end{figure}

\begin{fminipage}{\mylength}
\textbf{\underline{``Glued-trees'' problem}}
\vskip 6pt
\textbf{Input:}
A black-box implementing $O_{G_n}$ for a graph $G_n$ of the above form.

\textbf{Problem:} Output the label of the EXIT vertex.
\end{fminipage}

\begin{fminipage}{\mylength}
\textbf{\underline{Quantum Algorithms for the ``Glued-trees'' problem}}
\vskip 6pt
There is a continuous time quantum walk which starts at the ENTRANCE vertex (in this case $\ket{00 \ldots 0}$) and evolves according to the Hamiltonian defined by the adjacency matrix of the graph $G_n$ for an amount of time $t$ selected uniformly at random in $[0,\frac{n^4}{2 \epsilon}]$. Measuring will then yield the EXIT label with probability at least $\frac{1}{2n}(1 - \epsilon)$.

The authors show how to efficiently simulate this continuous time quantum walk using a universal quantum computer that makes a polynomial number of calls to $O_{G_n}$.
\end{fminipage}

\begin{fminipage}{\mylength}
\textbf{\underline{Classical Algorithms for the ``Glued-trees'' problem}}
\vskip 6pt
Any classical randomized algorithm must evaluate the black-box $O_{G_n}$ an exponential number of times in order to output the correct EXIT vertex label with non-negligible probability. More precisely, any classical algorithm that makes $2^{n/6}$ queries to $O_{G_n}$ can only find the EXIT with probability at most $4 \cdot 2^{-n/6}$.
\end{fminipage}

Another very interesting and recent problem for which a quantum walk algorithm has given the optimal algorithm is the problem of evaluating a NAND-tree (or AND-OR tree). The problem is nicely described by a binary tree of depth $n$ whose leaves are labelled by the integers $i \in \{1, 2, \ldots , 2^n \}$. The input is a black-box $O_{\X}$ that encodes a binary string $\X = X_1 X_2 \ldots X_N$, where $N = 2^n$. The $i$th leaf vertex is assigned value $X_i$, and the parent of any pair of vertices takes on the value which is the NAND of the value of its child vertices (the NAND of two input bits is $0$ if both inputs are $1$ and $1$ if either bit is $0$). Thus, given the assignment of values to the leaves of a binary tree, one can compute the values of the remaining vertices in the tree, including the root vertex. The value of the NAND tree for a given assignment is the value of the root vertex.

\begin{fminipage}{\mylength}
\textbf{\underline{NAND-tree evaluation}}
\vskip 6pt
\textbf{Input:}
A black-box $O_{\X}$ that encodes a binary string $\X = X_1 X_2 \ldots X_{N} \in \{0,1\}^N$, $N = 2^n$.

\textbf{Problem:}
Output the value of the binary NAND tree whose $i$th leaf has value $X_i$.
\end{fminipage}

\begin{fminipage}{\mylength}
\textbf{\underline{Classical Algorithms for NAND-tree evaluation}}
\vskip 6pt
The best known classical randomized algorithm uses $O(N^{0.753...})$ evaluations of the black-box, and it is also known that $\Omega(N^{0.753...})$ evaluations are required for any classical randomized algorithm. %\cite{SW86}
\end{fminipage}

Until recently, no quantum algorithm worked better.
\begin{fminipage}{\mylength}
\textbf{\underline{Quantum Algorithms for NAND-tree evaluation}}
\vskip 6pt
Farhi, Goldstone and Gutmann \cite{FGG07} showed a continuous time walk that could solve this it time $O(\sqrt{N})$ using a continuous version of the black-box, and it was subsequently showed that $O(N^{1/2 + \epsilon})$ queries to the discrete oracle suffice, for any real constant $\epsilon > 0$, and discrete walk versions of the algorithm and other generalizations were developed \cite{ACRSZ07, CCJY07}
\end{fminipage}

This was a very interesting breakthrough in solving a fundamental problem that had stumped quantum algorithms experts for a number of years.
The general idea is inspired from techniques in particle physics and scattering theory. They consider a graph formed by taking a binary tree and making two additions to it. Firstly, for each leaf vertex where $X_i = 1$, add another vertex and join it to that leaf. Secondly, attach the root vertex to the middle of a long line graph of length in $\Omega(\sqrt{N})$.
Then evolve the system according to the Hamiltonian equal to the adjacency matrix of this graph. Then one should start the system in a superposition of states on the left side of the line graph with phases defined so that if the NAND-tree were not attached, the ``wave packet'' would move from left to right along the line. If the packet gets reflected with non-negligible amplitude, then the NAND tree has value $1$, otherwise, if the packet gets mostly transmitted, then the NAND tree has value $0$. Thus one measures the system, and if one obtains a vertex to the left of the NAND-tree, one guesses ``1'', and if one obtains a vertex to the right of the NAND-tree, one guesses ``0''. This algorithm outputs the correct answer with high probability.

In a discrete query model, one can carefully simulate the continuous time walk \cite{CCJY07} (as discussed in section \ref{simulation}), or one can apply the results of Szegedy \cite{Sze04} to define a discrete-time coined walk with the same spectral properties \cite{ACRSZ07} and thus obtain a discrete query complexity of $N^{\frac{1}{2}+\epsilon}$ for any constant $\epsilon > 0$.

This algorithm has also been applied solve MIN-MAX trees with a similar improvement \cite{CGY08}. The NAND-tree and related problems are related to deciding the winner of two-player games.
Another very recent new class of quantum algorithms for evaluating a wider class of formulas, based on ``span'' programs, was developed in \cite{RS08}.

\section{Adiabatic algorithms} \label{adiabatic}

It is possible to encode the solution to a hard problem into the ground state of an efficiently simulatable Hamiltonian. For example, in order to try to solve $3$-SAT for a formula on $n$ Boolean variables, one could define a Hamiltonian
\[ H_1 = \sum_{\bf{x}\in \{0,1\}^n} f_{\Phi}(\bf{x}) \ket{\bf{x}}\bra{\bf{x}} \] where $f_{\Phi}(\bf{x})$ is the number of clauses of $\Phi$ that are violated by the assignment $\bf{x}$. Then one could try to define  algorithms to find such a ground state, such as quantum analogues of classical annealing
or other heuristics (e.g. \cite{Hog00, FGGS00}).

Adiabatic algorithms (also known as adiabatic optimization algorithms) are a new paradigm for quantum algorithms invented by Farhi, Goldstone, Gutmann and Sipser \cite{FGGS00}.
 The paradigm is based on the fact that, under the right conditions, a system that starts in the ground state of a Hamiltonian $H(0)$, will with high probability remain in the ground state of the Hamiltonian $H(t)$ of the system at a later time $t$, if the Hamiltonian of the system changes ``slowly enough'' from $H(0)$ to $H(t)$. This fact is called the adiabatic theorem (see e.g. \cite{JRS07}).

This theorem inspires the following algorithmic paradigm:
\begin{itemize}
\item
Convert your problem to generating the ground state of some easy-to-simulate Hamiltonian $H_1$.
\item
Initialize your quantum computer in an easy-to-prepare state $\ket{\psi_0}$ of an easy-to-simulate Hamiltonian $H_0$.
\item
On the quantum computer, simulate a time-dependent Hamiltonian $H(t) = (1-t/T) H_0 + t/T H_1$, for $t$ going from $0$ to $T$.
\end{itemize}

An important detail is how slowly to transition from $H_0$ to $H_1$ (in other words, how large $T$ should be).
This related to two important parameters. Let $\lambda_0(t)$ be the smallest eigenvalue of $H(t)$, and assume that the corresponding eigenspace is non-degenerate. Let $\lambda_1(t)$ be the second smallest eigenvalue of $H(t)$, and define $g(t) = \lambda_1(t) - \lambda_0(t)$ to be the gap between the two lowest eigenvalues.
The norm of the Hamiltonian is also relevant. This is to be expected, since one can effectively speed-up time by a factor of $s$ by just multiplying the Hamiltonian of the system by $s$. In any realistic implementation of the Hamiltonian one pays for such a speed-up by at least a factor of $s$ in some resource. For example, if we simulate the Hamiltonian using quantum circuits (e.g. as described in section \ref{simulation}), the overhead is a factor of $s^{1+o(1)}$ in the circuit depth, and thus there is no actual speed-up.
Furthermore, the norm of the derivatives of the Hamiltonian is also relevant.

There are a variety of theorems and claims in the literature proving, or arguing, that a value of $T$ polynomial in the operator norm of $\frac{d H(t)}{dt}$ (or even some higher derivative) and in the inverse of the minimum gap (i.e. the minimum $g(t)$, for $0 \leq t \leq T$), and in $\frac{1}{\delta}$, is sufficient in order to generate the final ground state with probability at least $1 - \delta$. The general folklore is that with the right assumptions the dependence on the minimum gap $g_{\min}$ is $\Omega(\frac{1}{g_{\min}^2})$, and one can find examples when this is the case, however more sophisticated descriptions of the dependence on $g(t)$ are known
(see e.g. \cite{JRS07}).

For example, in order to try to solve $3$-SAT for a formula on $n$ Boolean variables, one could define a Hamiltonian
\[ H_1 = \sum_{\bf{x}\in \{0,1\}^n} f_{\Phi}(\bf{x}) \ket{\bf{x}}\bra{\bf{x}} \] where $f_{\Phi}(\bf{x})$ is the number of clauses of $\Phi$ which are violated by the assignment $\bf{x}$. Assuming there is exactly one satisfying assignment $\bf{w}$, then $\ket{\bf w}\bra{\bf w}$ is the unique ground state of $H_1$. This algorithm has been studied numerically and analytically \cite{FGGLLP01, DMV01, RC02} and variations have been introduced as well to work around various lower bounds that were proved \cite{FGG02}.
Unfortunately, it is not known what the worst-case complexity is for these algorithms on such NP-hard problems, since it has proved very hard to provide rigorous or even convincing heuristic bounds on the minimum gap for such problems of interest. It is widely believed that these algorithms will not solve an NP-hard problem in polynomial time, partly because it is believed that no quantum algorithm can do this.

A slight generalization of adiabatic algorithms, called adiabatic computation (where the final Hamiltonian does not need to be diagonal in the computational basis) was shown to be polynomially equivalent to general quantum computation \cite{ADKLLR04}.

\section{Topological algorithms} \label{topological}

The standard models of quantum computation (e.g. quantum Turing machine, quantum acyclic circuits) are known to be equivalent in power (up to a polynomial factor), and the quantum Strong Church-Turing thesis states that any realistic model of computation can be efficiently simulated by such a quantum computer. If this were not the case, then one should seek to define a stronger model of computation that encapsulates the full computational power that the laws of physics offer.

Freedman \cite{Fre98} proposed defining a computing model based on topological quantum field theories. The main objective was that such a computer might naturally solve an NP-hard or \#P-hard topological problem, in particular, evaluating the Jones polynomial at certain points. A natural family of such topological quantum field theory computers was shown to be equivalent in power to the standard model of quantum computation, thus such a new model of computation would not provide additional computational power, but it was hoped that this new paradigm might inspire new quantum algorithms. In fact, it has been shown that ``topological'' algorithms can approximate the value of the Jones polynomial at certain points more efficiently than any known classical algorithm \cite{FKW02, AJL06}. The known approximations are not good enough to solve an NP-hard problem. Several other generalizations and related problems and algorithms have been found recently \cite{LK06, WY08, AAEL07}. We will briefly sketch some of the definitions, results and techniques.

A {\it knot} is a closed non-intersecting curve embedded in $\bf{R}^3$, usually represented via a knot diagram, which is a projection of the knot into the plane with additional information at each cross-over to indicate which strand goes over and which goes under. Two knots are considered equivalent if one can be manipulated into the other by an isotopy (i.e. by a  transformations one could make to an actual knot that can be moved and stretched but not broken or passed through itself).
A link is a collection of non-intersecting knots embedded in $\bf{R}^3$. They can be represented by similar diagrams, and there is a similar notion of equivalence.

The Jones polynomial of a link $L$ is a polynomial $V_L(t)$ that is a link invariant; in other words it has the property that if two links $L_1$ and $L_2$ are equivalent, then $V_{L_1}(t) = V_{L_2}(t)$.
Computing the Jones polynomial is in general \#P-hard for all but a finite number of values of $t$.
In particular, it is \#P-hard to evaluate the Jones polynomial exactly at any primitive $r$th root of unity for any integer $r \geq 5$. However, certain approximations of these values are not known to be \#P-hard.

The Jones polynomial is a special case of the Tutte polynomial of a planar graph. For a planar graph $G = (V,E)$, with weights edge weights $\mathbf{v} = \{v_e \mid e \in E\}$ the multivariate Tutte polynomial is defined as
\[ Z_G(q; v_{e_1}, v_{e_2}, \ldots ) = \sum_{A \subseteq E} q^{k(A)} \prod_{e \in A} v_e \] where $q$ is another variable and $k(A)$ is the number of connected components in the subgraph $(V,A)$. The standard Tutte polynomial $T_G(x,y)$ is obtained by setting $v_e = v$ for all $e \in E$, $x = 1 + q/v$ and $y = 1+v$.
 Connections with physics are discussed, for example, in  \cite{Kau01, Wel93, Sok05}.

Here we briefly sketch a specific instance of such a problem, and the approach of \cite{AAEL07} taken to solve this problem on a quantum computer.

Firstly, for any planar graph $G$, one can efficiently find its {\it medial graph} $L_G$, which is a $4$-regular planar graph which can be drawn from a planar embedding of $G$ as follows. Draw a new vertex with weight $u_i$ in the middle of each edge that had label $v_i$. For each new vertex, on each side of the original edge on which the vertex is placed, draw a new edge going in the clockwise direction joining the new vertex to the next new vertex encountered along the face of the original graph. Do the same in the counter-clockwise direction, and remove all the original edges and vertices.

From this medial graph, one can define another polynomial called the {\it Kauffman bracket}, denoted $\langle L_G \rangle (d, u_1, u_2, \ldots )$, that satisfies $\langle L_G \rangle (d, u_1, u_2, \ldots) = d^{-|V|} Z_G(d^2; d u_1, d u_2 \ldots )$.

\begin{fminipage}{\mylength}
\textbf{\underline{Additive approximation of the multivariate Tutte polynomial}}
\textbf{\underline{for a planar graph}}

\textbf{Input:}
A description of a planar graph $G = (V,E)$.

Complex valued weights $v_1, v_2, \ldots, v_m$ corresponding to the edges $e_1, e_2, \ldots, e_m \in E$ of $G$.

A complex number $q$.

\textbf{Problem:}
Output an  approximation of $Z_G(q; v_{1}, v_{2}, \ldots, v_m)$.
\end{fminipage}

\begin{fminipage}{\mylength}
\textbf{\underline{Quantum Algorithms for Approximating the  Tutte Polynomial}}
\vskip 6pt
Aharonov et al. give a quantum algorithm that solves the above problem in time polynomial in $n$ with an additive approximate they denote by $\Delta_{alg}/\poly(m)$, which is described below.
\end{fminipage}

The value $\Delta_{alg}$ depends on the embedding of the graph. The results are hard to compare to what his known about classical algorithms for this problem (see \cite{BFLW05} for a discussion), but there are special cases that are BQP-hard.

\begin{fminipage}{\mylength}
\textbf{\underline{Classical Algorithms for Approximating the Tutte Polynomial}}
\vskip 6pt
It was shown \cite{AAEL07} that for certain ranges of parameter choices, the approximations given by the quantum algorithms are BQP-hard, and thus we don't expect a classical algorithm to be able to provide as good of an approximation unless classical computers can efficiently simulate quantum computers.
\end{fminipage}

\subsection{Sketch of the structure of the algorithm}

We only have room to give a broad overview of the algorithm. One of the main points is to emphasize that this algorithm looks nothing any of the other algorithms discussed in the previous sections.

At a very high level, the idea is that these medial graphs $T_G$ can be represented by a linear operation $Q_G$ such that $L_G(d, u_1, u_2, \ldots ) = \bra{1} Q_G \ket{1}$.
The quantum algorithm approximates the inner product between $\ket{1}$ and $Q_G \ket{1}$, and therefore gives an approximation to the Kauffman bracket for $G$ and thus the generalized Tutte polynomial for $G$.

The medial graph $L_G$ will be represented as a product of basic elements $\mathcal{T}_i$ from a generalized Temperley-Lieb algebra. These basic elements $\mathcal{T}_i$ will be represented by simple linear transformations on finite dimensional Hilbert spaces, which can be implemented on a quantum computer.
Below we briefly sketch this decomposition and correspondence.

One can easily draw the medial graph $L_G$ in the plane so that one can slice it with horizontal lines so that in between each consecutive horizontal line there is a diagram with only one of the following: a crossing of two lines, a ``cap'' or a ``cup''.
%, as illustrated in figure \ref{cap+cup}.
One can think of the gluing together of these adjacent diagrams $\mathcal{T}_i$ to form the graph $L_G$ as a product operation. We will sketch how to map each $\mathcal{T}_i$ to a linear operation $\rho(\mathcal{T}_i)$ acting on a finite dimensional Hilbert space.

The state space that the operation $\rho(\mathcal{T}_i)$ will act on is the set of finite walks starting at vertex $1$ on the infinite graph $G$ with vertices labelled by the non-negative integers and edges $\{i,i+1\}$ for all $i \geq 0$. For example, the walk $1-2-3-2-3$ is a walk of length $4$ from $1$ to $3$.
The linear transformation $\rho(\mathcal{T}_i)$ maps a walk $w_1$ to a linear combination of walks that are ``compatible'' with $\mathcal{T}_i$ (we won't explain here the details of defining this linear transformation).

In order to apply this technique to a diagram with a crossing,
one eliminates the crossing by replacing it with two non-intersecting lines. This replacement can be done in two different ways, and one can represent the diagram with a crossing as a formal linear combination of these two diagrams one gets by replacing a crossing at a vertex (say with label $u$)  with two non-intersecting lines, where one of the two links gets the coefficient $u$ (by a simple rule that we don't have room to explain). %This is illustrated in figure \ref{medial2}.
We can then apply the construction
%illustrated in figure \ref{tangle-path}
to each of these new diagrams, and combine them linearly to get the linear transformation corresponding to the diagram with the crossing.

These linear transformations are not necessarily unitary (they were unitary in the earlier work on the Jones polynomial, and other related work, construct unitary representations), however the authors show how one can use ancilla qubits and unitaries to implement non-unitary transformations and approximate the desired inner product using the ``Hadamard test'' (see Section \ref{quantumtasks}).

\section{Quantum algorithms for quantum tasks} \label{quantumtasks}

At present, when we think of quantum algorithms, we usually think of starting with a classical input, running some quantum algorithm, or series of quantum algorithms, with some classical post-processing in order to get a classical output. There might be some quantum sub-routines that have been analyzed, but the main goal in mind is to solve a classical problem.

For example, quantum error correction (see \cite{Kem07} for a recent survey)
can be thought of as an algorithm having a quantum input and a quantum output.
There are many algorithms for transferring a qubit of information through a network of qubits under some physical constraints. We might develop a quantum cryptographic infrastructure where objects like money and signatures \cite{GC01} are quantum states that need to be maintained and manipulated as quantum states for long periods of time.

Several of the algorithms described in the previous sections have versions which have a quantum input or a quantum output or both. For example, the amplitude amplification algorithm can be rephrased in terms of finding a quantum state $\ket{\phi}$ given a black-box $U_{\phi}$ that recognized $\ket{\phi}$ by mapping $\ket{\phi} \mapsto - \ket{\phi}$ and acting as the identity on all states orthogonal to $\ket{\phi}$. The end result of such a quantum search is the quantum state $\ket{\phi}$. Amplitude estimation is estimating a transition probability of a unitary operator.

The topological algorithms require as a basic subroutine a quantum algorithm for approximating the inner product of $\ket{0}$ and $U \ket{00 \ldots 0}$, which the authors call the ``Hadamard test''. The algorithm consists of using a controlled-$U$ operation to create the state $\frac{1}{\sqrt{2}} \ket{0} \ket{00 \ldots 0} + \frac{1}{\sqrt{2}} \ket{1} U \ket{00 \ldots 0}$. Note that if we apply the Hadamard gate to the first qubit, we get
\[ \sqrt{\frac{1+ Re\bra{00 \ldots 0} U \ket{00 \ldots 0}}{2}} \ket{0} \ket{\psi_0} +  \sqrt{\frac{1- Re\bra{00 \ldots 0} U \ket{00 \ldots 0}}{2}} \ket{1} \ket{\psi_1} \]
for normalized states $\ket{\psi_0}$ and $\ket{\psi_1}$. We can thus estimate the real part of the $\bra{00 \ldots 0} U \ket{00 \ldots 0}$ by repeating several times, or applying the quadratically more efficient amplitude estimation algorithm algorithm described earlier (the goal is a superpolynomial speed-up, so a quadratic improvement is not substantial in this case).  We can also estimate the complex part similarly.

Another example is the {\it coset orbit problem} \cite{FIMSS03} mentioned in section \ref{hsp}. The input to this problem consists of two quantum states $\ket{\phi_0}$ and $\ket{\phi_1}$ from a set $\Gamma$ of mutually orthogonal states and black-boxes for implementing the action of a group $G$ on the set $\Gamma$. For a state $\ket{\phi}$, let $\ket{u \cdot \phi}$ denote the state resulting from the action of $u \in G$ on $\ket{\phi}$, and let $G_{\ket{\phi}}$ denote the subgroup of $G$ that stabilizes the state $\ket{\phi}$ (i.e. the set of  $u \in G$ such that $\ket{u \cdot \phi} = \ket{\phi}$). The question is whether there exists a $u \in G$ such that $\ket{u \cdot \phi_1} = \ket{\phi_0}$.
If the answer is ``yes'', then the set of $u$ satisfying $\ket{ u \cdot \phi_1} = \ket{\phi_0}$ is a left coset of $G_{\ket{\phi_1}}$. Thus, the algorithm should output a coset representative $u$ along with $O(\log n)$ generators of $G_{\ket{\phi_1}}$.
The solution to this problem was an important part of the solution to the hidden translation problem. Several other quantum algorithms have such ``quantum'' sub-routines.

Other examples of quantum tasks include quantum data compression which was known to be information theoretically possible, and efficient quantum circuits for performing it were also developed \cite{BHL06}. Another example is entanglement concentration and distillation.

Researchers have also developed quantum algorithms for implement some natural basis changes, which have quantum inputs and quantum outputs. For example, the Clebsch-Gordan transformation \cite{BCH06} or other transformations (e.g. \cite{Hoy97, Ett00}).

We've only listed a few examples here. In the future, as quantum technologies, in particular quantum computers and reliable quantum memories, develop, quantum states and their manipulation will become an end in themselves.

\section{Future directions} \label{future}

Roughly 10 years ago, many people said that there were essentially only two quantum algorithms. One serious omission was the simulation of quantum mechanical systems, which was in fact Feynman's initial motivation for quantum computers. Apart from this omission, it was true that researchers were developing a better and deeper understanding of the algorithms of Shor and Grover, analyzing them in different ways, generalizing them, and applying them in non-trivial ways. It would have been feasible to write a reasonably sized survey of all the known quantum algorithms with substantial details included. In addition to this important and non-trivial work, researchers were looking hard for ``new'' approaches, and fresh ideas like quantum walks, and topological algorithms, were being investigated, as well as continued work on the non-Abelian version of the hidden subgroup problem. The whole endeavour of finding new quantum algorithms was very hard and often frustrating. Fortunately, in the last 10 years, there have been many non-trivial developments, enough to make the writing of a full survey of quantum algorithms in a reasonable number of pages impossible. Some directions in which future progress might be made are listed below.

\begin{itemize}
\item
The complexity of the non-Abelian hidden subgroup problem will hopefully be better understood. This includes addressing the question: Does there exist a quantum polynomial time algorithm for solving the graph isomorphism problem?
Of course, a proof that no such quantum algorithm exists would imply that $P \neq PSPACE$. So, more likely, we might develop a strong confidence that no such algorithm exists, in which case this can form the basis of quantum  computationally secure cryptography \cite{MRV07}.

\item
There are many examples where amplitude amplification was used to speed-up a classical algorithm in a non-trivial way. That is, in a way that was more than just treating a classical algorithm as a guessing algorithm $A$ and applying amplitude amplification to it. There are likely countless other algorithms which can be improved in non-trivial ways using amplitude amplification.

\item
The quantum walk paradigm for quantum algorithms emerged roughly 10 years ago, and has recently been applied to find the optimal black-box algorithm for several problems, and has become a standard approach for developing quantum algorithms. Some of these black-box problems are fairly natural, and the black-boxes can be substituted with circuits for actual functions of interest. For example, collision finding can be applied to find collisions in actual hash functions used in cryptography. We will hopefully see more instances where black-box algorithm can be applied to solve an problem without a black-box, or where there is no black-box in the first place.

\item
In addition to the development of new quantum walk algorithms, we will hopefully have a more elegant and unified general theory of quantum walks that unites continuous and discrete walks, coined and non-coined walks, and quantum and classical walks.

\item
The adiabatic algorithm paradigm has not reached the level of success of quantum walks, partly because it is hard to analyze the worst case complexity of the algorithms. To date there is no adiabatic algorithm with a proof that it works more than quadratically faster than the best known classical algorithm. Can we do better with an adiabatic algorithm?

If and when we have large-scale quantum computers, we will be able to just test these algorithms to see if indeed they do have the conjectured running times on instances of interesting size.

\item
The topological algorithms have received limited attention to date. This is partly because the initial work in this field was largely inaccessible to researchers without substantial familiarity with topological quantum field theory and related areas of mathematics. The more recent work summarized in this paper and other recent papers is a sign that this approach could mature into a fruitful paradigm for developing new important quantum algorithms.

\item
The paradigm of measurement based computation (see e.g. \cite{Joz06} for an introduction) has been to date mostly focussed on its utility as a paradigm for possibly implementing a scalable fault-tolerant quantum computer. We might see the development of algorithms directly in this paradigm. Similarly for globally controlled architectures.
%MM todo

\item
There is also a growing group of researchers looking at the computational complexity of various computational problems in physics, in particular of simulating certain Hamiltonian systems, often coming from condensed matter physics.
Much of the work has been complexity theoretic, such as proving the $QMA$-hardness of computing ground states of certain Hamiltonians (e.g. \cite{AGIK07}). Other work has focussed on understanding which quantum systems can be simulated efficiently on a classical computer. This work should lead to the definition of some simulation problems that are not known to be in $BPP$, nor believed to be $NP$-hard or $QMA$-hard, and thus might be good candidates for a quantum algorithm. There has been a language and culture barrier between physicists and theoretical computer scientists when it comes to discussing such problems. However, it is slowly breaking down, as more physicists are becoming familiar with algorithms and complexity, and more quantum computer scientists are becoming familiar with language and notions from physics. This will hopefully lead to more quantum algorithms for computational problems in physics, and new algorithmic primitives that can be applied to a wider range of problems.

\end{itemize}

In summary, as daunting as it is to write a survey of quantum algorithms at this time, it will be a much harder task in another 10 years. Furthermore, in another 10 years we will hopefully have a better idea of when we might expect to see quantum computers large enough to solve problems faster than the best available classical computers.

%%%%%%%%%%% The bibliography starts:

\end{document}